\documentclass[6pt,onecolumn]{article}
\usepackage[english]{babel}

\usepackage{amsmath}
\usepackage{amsfonts}
\usepackage{amssymb}
\usepackage{graphicx}
\usepackage[affil-it]{authblk}
\usepackage{hyperref}
\usepackage{color}
\usepackage{caption}
\usepackage{verbatim}
\usepackage[normalem]{ulem}
\usepackage{subfigure}
\usepackage{slashed}
\usepackage{cancel}
\usepackage{braket}
\usepackage{multicol}
\usepackage{tcolorbox}
\tcbuselibrary{theorems}
\usepackage{bm}
\usepackage{enumitem}

\hypersetup{linktocpage=true}

\usepackage{tikz} 
\usepackage{tikz-feynman}

\tikzfeynmanset{compat=1.0.0}

\usetikzlibrary{arrows,shapes}
\usetikzlibrary{trees}
\usetikzlibrary{matrix,arrows} 				
\usetikzlibrary{positioning}				
\usetikzlibrary{calc,through}				
\usetikzlibrary{decorations.pathreplacing}  
\usepackage{pgffor}							

\usetikzlibrary{decorations.pathmorphing}	
\usetikzlibrary{decorations.markings}
\tikzset{
    vector/.style={decorate, decoration={snake}, draw},
	provector/.style={decorate, decoration={snake,amplitude=2.5pt}, draw},
	antivector/.style={decorate, decoration={snake,amplitude=-2.5pt}, draw},
    fermion/.style={draw=black, postaction={decorate},
        decoration={markings,mark=at position .55 with {\arrow[draw=black]{>}}}},
    fermionbar/.style={draw=black, postaction={decorate},
        decoration={markings,mark=at position .55 with {\arrow[draw=black]{<}}}},
    fermionnoarrow/.style={draw=black},
    gluon/.style={decorate, draw=black,
        decoration={coil,amplitude=4pt, segment length=5pt}},
    scalar/.style={dashed,draw=black, postaction={decorate},
        decoration={markings,mark=at position .55 with {\arrow[draw=black]{>}}}},
    scalarbar/.style={dashed,draw=black, postaction={decorate},
        decoration={markings,mark=at position .55 with {\arrow[draw=black]{<}}}},
    scalarnoarrow/.style={dashed,draw=black},
    electron/.style={draw=black, postaction={decorate},
        decoration={markings,mark=at position .55 with {\arrow[draw=black]{>}}}},
	bigvector/.style={decorate, decoration={snake,amplitude=4pt}, draw},
    line/.style={draw=black},
}\usetikzlibrary{decorations.markings}

\title{Bouncing pNGB Dark Matter via a Fermion Dark Matter}
\author[1,2]{Basti\'an D\'iaz S\'aez\thanks{bastian.diaz.s@usach.cl}}
\author[3,4]{Patricio Escalona Contreras\thanks{patricio.escalona@sansano.usm.cl}}
\affil[1]{Department of Physics, University of Santiago of Chile, Casilla 307, Santiago, Chile}
\affil[2]{II. Institut für Theoretische Physik, Universität Hamburg, 22761 Hamburg, Germany}
\affil[3]{Department of physics, Federico Santa Maria Technical University, Avenida España 1680, Valparaíso, Chile}
\affil[4]{Millennium Institute for Subatomic Physics at High-Energy Frontier
(SAPHIR), Fern\'andez Concha 700, Santiago, Chile}

\begin{document}
\maketitle

\begin{abstract}
In addition to the Standard Model, the introduction of a singlet complex scalar field that acquires vacuum expectation value may give rise to a cosmologically stable pseudo-Nambu-Goldstone boson (pNGB), a suitable dark matter (DM) candidate. This work extends this scenario
by including a second cosmologically stable particle: a fermion singlet. The pNGB and the new fermion can be regarded as DM candidates simultaneously, both interacting with the Standard Model through Higgs portals via two non-degenerate Higgs bosons. We explore the thermal freeze-out of this scenario, with particular emphasis on the increasing yield of the pNGB before it completely decouples (recently called \textit{Bouncing DM}). We test the model under collider bounds, relic abundance, and direct detection, and we explore some indirect detection observables today. 

\end{abstract}

\tableofcontents
\section{Introduction}
The dark matter (DM) field is an active area of research comprehending new extensions to the Standard Model (SM) of particle physics. One simple well-motivated scenario is the global $U(1)$ complex scalar extension to the SM, which after the spontaneous and explicit symmetry breaking gives rise to a pseudo-Nambu-Goldstone boson (pNGB) and a second Higgs boson \cite{Barger:2008jx, Barger:2010yn, Gross:2017dan}. This pNGB, being cosmologically stable, has been studied as a thermal relic in a standard freeze-out, presenting viable DM masses in the ballpark of the electroweak scale. A second Higgs may be relevant in the explanations of anomalies seen at LEP and LHC \cite{Cline:2019okt}.   

On the other hand, the question of whether the DM genesis was thermal or not has given rise to interesting possibilities. For instance, non-thermal DM, such as feebly-interacting massive particles (FIMPS) \cite{Feng:2003xh}, or DM subject to an exponential growth \cite{Bringmann:2021tjr}, evolve increasing their yields from a negligible initial abundance until Hubble expansion dominates and the relic abundance is set. Analogously, could thermal relics present a feature like this growing yield in the early universe? Recently, it has been suggested that in specific scenarios, the yield of a DM particle could have been undergoing a different pattern of decoupling, with a period of exponential growth before complete freeze-out. This idea has been called \textit{bouncing dark matter} \cite{Puetter:2022ucx}. Although this bouncing effect in the yield of some species has been already observed in the context of thermal DM, see e.g. \cite{Katz:2020ywn, Ho:2022erb, Ho:2022tbw, Ghosh:2023dhj}, the connection of this feature with indirect detection has only been worked out in \cite{Puetter:2022ucx}.

In this work, we study a multi-component DM scenario in which one of the stable relics undergoes this bouncing effect. The model consists of the introduction of two gauge singlets: a fermion and a complex scalar field, both transforming under a global symmetry\footnote{For related model building constructions see \cite{Weinberg:2013kea, Garcia-Cely:2013wda, Arcadi:2017kky, Yaguna:2021rds}.}. After spontaneous symmetry breaking (SSB), the singlet scalar gets vacuum expectation value, giving rise to a second Higgs boson which will mix with the Higgs excitation of the SM Higgs doublet, and a Nambu-Goldstone boson $\chi$. Imposing a $U(1)$ soft-breaking term in the scalar sector, $\chi$ becomes a cosmological stable pNGB if its mass is lower than twice the fermion singlet mass. In this way, the model presents two DM components that communicate to the SM through two Higgs-like particles. 

We study the thermal freeze-out of both stable relics, with special emphasis on the details of the bouncing effect undergone by the yield of the pNGB, and also some prospects of indirect detection. Finally, as the minimal scenario of the pNGB as the only source of DM allows masses above 50 GeV \cite{Gross:2017dan}, unless resonance annihilation effects are present, in the present multi-component scenario we explore the viability to have pNGB DM masses below 50 GeV. 

The paper is organized in the following way. In Sec.~\ref{model} we introduce the model. In Sec~\ref{relicabundance} we present the Boltzmann equations for the system, with a detailed analysis of the relic abundance of both components, highlighting the bouncing of the pNGB yield, and some average cross sections relevant for indirect detection. In Sec.~\ref{pheno} we present the phenomenology of the model for two-DM components, imposing relevant constraints, and studying some indirect signals. Finally, in Sec~\ref{conclusions} we discuss and conclude our work.  

\section{Model}\label{model}
Aside from the SM particle content, we add two SM singlets, a Dirac fermion $\psi$ (for the Majorana case the construction follows equivalently, e.g. \cite{Garcia-Cely:2013wda}) and a complex scalar $S$. Decomposing $\psi = \psi_L + \psi_R$, we impose a chiral approximate global symmetry $U(1)_V\times U(1)_A$, with the new fields transforming as \cite{Pokorski:2021qgt}
\begin{eqnarray}\label{va}
 U(1)_V&:&\quad S \rightarrow S ,\quad\quad \psi_L\rightarrow e^{i\frac{\beta_V}{2}}\psi_L , \quad\quad \psi_R\rightarrow e^{i\frac{\beta_V}{2}}\psi_R  ,\\
 U(1)_A&:&\quad S \rightarrow e^{i\beta_A} S ,\quad\quad \psi_L\rightarrow e^{i\frac{\beta_A}{2}}\psi_L , \quad\quad \psi_R\rightarrow e^{-i\frac{\beta_A}{2}}\psi_R ,
\end{eqnarray}
with $\beta_{V,A}$ arbitrary constants. More concisely, the spinor under $U(1)_V\times U(1)_A$ transform as $\psi\rightarrow e^{i(\beta_V + \gamma^5 \beta_A)}\psi$. This symmetry and particle content give rise to the Lagrangian:
\begin{eqnarray}\label{lag1}
 \mathcal{L}_{BSM} = \bar{\psi}i\slashed{\partial}\psi + (\partial_\mu S)^{\dagger}\partial^{\mu}S - g_{\psi}\bar{\psi}_L\psi_R S - g_{\psi}^*\bar{\psi}_R\psi_L S^{\dagger} - V(H , S),
\end{eqnarray}
with the potential given by
\begin{eqnarray}\label{pot1}
 V(H , S) = -\frac{\mu_H^2}{2}|H|^2 - \frac{\mu_S^2}{2}\vert S \vert^2  + \frac{\lambda_H}{2}\vert H \vert^4 + \frac{\lambda_S}{2}\vert S \vert^4  + \lambda_{HS}\vert H \vert^2 \vert S \vert^2 + V_\text{soft}.
\end{eqnarray}
Here $H$ is the Higgs field, and the $U(1)_A$ soft breaking term is given by
\begin{eqnarray}\label{soft}
 V_\text{soft} = -\frac{m_\chi^2}{2}\left(S^2 + S^{*2}\right),
\end{eqnarray}
where we have assumed that a subgroup $Z_2$ of $U(1)_A$ remains unbroken (i.e. $\beta_A = \pi$), in such a way that only soft-breaking terms with even powers of $S$ are allowed. 
The couplings $g_\psi$ and $m_\chi^{2}$ can be made real by field redefinitions. Note that the global symmetry does not allow a mass term for the fermion singlet, but it will be generated through the spontaneous symmetry breaking of $S$. Analogously to the global $U(1)$ complex scalar model \cite{Gross:2017dan}, the Lagrangian is invariant under a \textit{CP-symmetry}: $\psi_{L,R}\rightarrow \psi_{R,L}$ and $S\rightarrow S^*$. Notice that $Z_2$ symmetry in which $S\rightarrow -S$ is broken by the vacuum $v_s$ \cite{Azevedo:2018oxv}. 

The stationary point conditions at $(h,s) = (0,0)$ are
\begin{eqnarray}
 \mu_H^2 &=& \lambda_H v_h^2 + \lambda_{HS}v_s^2 , \\
 \mu_S^2 &=& \lambda_{HS} v_h^2 + \lambda_{S}v_s^2 - m_\chi^2 .
\end{eqnarray}
As $\mu_S^2 < 0$, $S$ triggers the SSB, making $U(1)_A$ symmetry is spontaneously broken. Explicitly, choosing the Hermitian basis, $S = (s' + i\chi)/\sqrt{2}$,  the minimum is acquired when ${s'}^2 + \chi^2 = v_s^2 > 0$, with $v_s$ the vacuum expectation value of the complex field. Without loss of generality, we set $\braket{s'} = v_s$ and $\braket{\chi} = 0$. Noting that $\bar{\psi}_L\psi_R + \bar{\psi}_R\psi_L = \bar{\psi}\psi$ and $\bar{\psi}_L\psi_R - \bar{\psi}_R\psi_L = \bar{\psi}\gamma^5\psi$, the interactions around the new vacuum after the SSB of $U(1)_A$ are given by
\begin{eqnarray}\label{lag2}
 \mathcal{L} = - \frac{g_{\psi}v_s}{\sqrt{2}}\bar{\psi}\psi\left(1 + \frac{s}{v_s}\right) - \frac{g_{\psi}}{\sqrt{2}}i\bar{\psi}\gamma^5\psi \chi - V(H , s , \chi). \nonumber
\end{eqnarray}
Note that the fermion has acquired a mass $m_\psi \equiv g_\psi v_s/\sqrt{2}$, a scalar interaction with $s$, and a pseudo-scalar interaction with $\chi$.
After electroweak symmetry breaking (EWSB), the Higgs field mixes with $s$, which introduces a mixing angle $\theta$ which satisfies
\begin{eqnarray}\label{thetamix}
 \tan 2\theta = \frac{2\lambda_{HS}v_hv_s}{\lambda_S v_s^2 - \lambda_H v_h^2},
\end{eqnarray}
with $v_h = 246$ GeV, $h = \cos\theta h_1 + \sin\theta h_2$ and $s = -\sin\theta h_1 + \cos\theta h_2$, and the masses of the scalars are
\begin{eqnarray}
 m_{h_1,h_2}^2 = \frac{1}{2}\left(\lambda_S v_s^2 + \lambda_H v_h^2 \mp \frac{\lambda_S v_s^2 - \lambda_H v_h^2}{\cos 2\theta}\right),
\end{eqnarray}
where we identify $h_1$ with the 125 GeV Higgs boson. From this last relation, we see that 
\begin{eqnarray}\label{quartic3}
 \lambda_H &=& \frac{1}{2v_h^2}\left[m_{h_1}^2 + m_{h_2}^2 + \cos 2\theta(m_{h_1}^2 - m_{h_2}^2)\right], \\
 \lambda_S &=& \frac{1}{2v_s^2}\left[m_{h_1}^2 + m_{h_2}^2 - \cos 2\theta(m_{h_1}^2 - m_{h_2}^2)\right].
\end{eqnarray}
In the physical basis, the interaction of the singlet fermion with the scalars is given by
\begin{eqnarray}\label{masterlag}
\mathcal{L} \supset - \frac{g_{\psi}}{\sqrt{2}}\bar{\psi}\psi(-h_1 \sin\theta + h_2\cos\theta) - \frac{g_{\psi}}{\sqrt{2}}i\bar{\psi}\gamma^5\psi \chi - V(h_1 , h_2 , \chi),
\end{eqnarray}
with the potential $V(h_1 , h_2 , \chi)$ explicitly detailed in the App.~\ref{potentialap}. The free parameters of the model are three masses and two couplings:
\begin{eqnarray}\label{parameters}
 \{m_\psi , m_\chi, m_{h_2} ,  g_\psi , \theta \}.
\end{eqnarray}
The model described by eq.~\ref{masterlag} may present one or two DM candidates, depending on the mass hierarchy between $\psi$ and $\chi$. If $m_\chi \geq 2m_\psi$, the fermion is the only stable field, and for $m_\chi < 2m_\psi$ the pNGB becomes stable, then the model presents two DM candidates (for a more detailed study of the stability of the pNGB, see \cite{DiazSaez:2021pmg}. Also see \cite{Belyaev:2022qnf}). In this work, we consider the latter case.
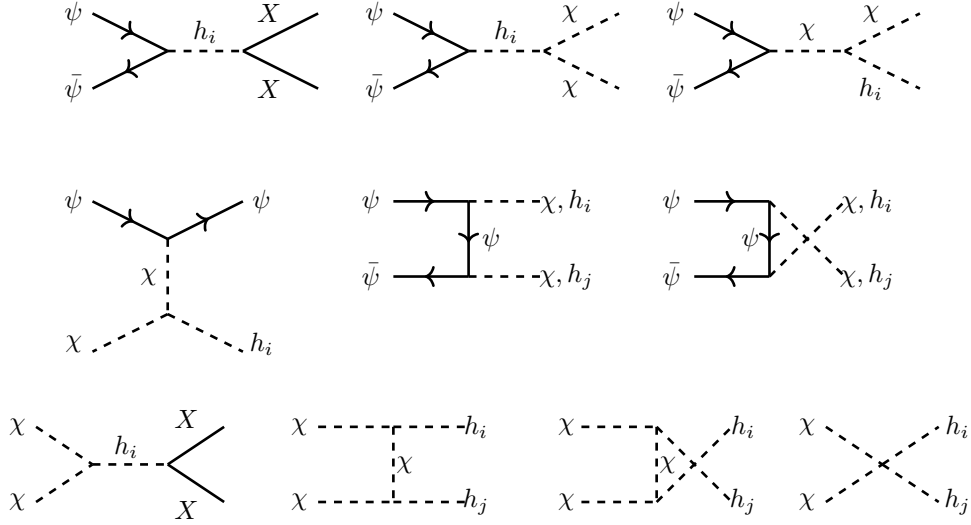
\begin{figure}[t!]
\centering
\begin{tikzpicture}[line width=1.0 pt, scale=0.5]
\begin{scope}[shift={(0,5)}]
	\draw[fermion](-3,1) -- (-1,0);
	\draw[fermionbar](-3,-1) -- (-1,0);
	\draw[scalarnoarrow](-1,0) -- (1,0);
	\draw[line](1,0) -- (3,1);
	\draw[line](1,0) -- (3,-1);
   \node at (-3.5,1.0) {$\psi$};
	\node at (-3.5,-1.0) {$\bar{\psi}$};
   \node at (0,0.5) {$h_i$};
	\node at (1.7,1) {$X$};
   \node at (1.7,-1) {$X$};
\end{scope}

\begin{scope}[shift={(8,5)}]
	\draw[fermion](-3,1) -- (-1,0);
	\draw[fermionbar](-3,-1) -- (-1,0);
	\draw[scalarnoarrow](-1,0) -- (1,0);
	\draw[scalarnoarrow](1,0) -- (3,1);
	\draw[scalarnoarrow](1,0) -- (3,-1);
   \node at (-3.5,1.0) {$\psi$};
	\node at (-3.5,-1.0) {$\bar{\psi}$};
   \node at (0,0.5) {$h_i$};
	\node at (1.7,1) {$\chi$};
   \node at (1.7,-1) {$\chi$};
\end{scope}

\begin{scope}[shift={(16,5)}]
	\draw[fermion](-3,1) -- (-1,0);
	\draw[fermionbar](-3,-1) -- (-1,0);
	\draw[scalarnoarrow](-1,0) -- (1,0);
	\draw[scalarnoarrow](1,0) -- (3,1);
	\draw[scalarnoarrow](1,0) -- (3,-1);
   \node at (-3.5,1.0) {$\psi$};
	\node at (-3.5,-1.0) {$\bar{\psi}$};
   \node at (0,0.5) {$\chi$};
	\node at (1.7,1) {$\chi$};
   \node at (1.7,-1) {$h_i$};
\end{scope}

\begin{scope}[shift={(0,0)}]
	\draw[fermion](-3,1) -- (-1,0);
	\draw[fermion](-1,0) -- (1,1);
	\draw[scalarnoarrow](-1,0) -- (-1,-2);
	\draw[scalarnoarrow](-1,-2) -- (-3,-3);
	\draw[scalarnoarrow](-1,-2) -- (1,-3);
   \node at (-3.5,1.0) {$\psi$};
	\node at (1.5,1.0) {$\psi$};
   \node at (-1.5,-1) {$\chi$};
	\node at (-3.5,-2.8) {$\chi$};
   \node at (1.5,-2.8) {$h_i$};
\end{scope}

\begin{scope}[shift={(8,0)}]
 \draw[fermion](-3,1) -- (-1,1);
	\draw[fermionbar](-3,-1) -- (-1,-1);
	\draw[fermion](-1,1) -- (-1,-1);
	\draw[scalarnoarrow](-1,1) -- (1,1);
	\draw[scalarnoarrow](-1,-1) -- (1,-1);
    \node at (-3.6,1.0) {$\psi$};
	\node at (-3.6,-1.0) {$\bar\psi$};
    \node at (-0.4,0) {$\psi$};
	\node at (1.6,1) {$\chi,h_i$};
    \node at (1.6,-1) {$\chi,h_j$};
\end{scope}

\begin{scope}[shift={(16,0)}]
	\draw[fermion](-3,1) -- (-1,1);
	\draw[fermionbar](-3,-1) -- (-1,-1);
	\draw[fermion](-1,1) -- (-1,-1);
	\draw[scalarnoarrow](-1,1) -- (1,-1);
	\draw[scalarnoarrow](-1,-1) -- (1,1);
    \node at (-3.6,1.0) {$\psi$};
	\node at (-3.6,-1.0) {$\bar\psi$};
    \node at (-1.5,0) {$\psi$};
	\node at (1.6,1) {$\chi, h_i$};
    \node at (1.6,-1) {$\chi, h_j$};
\end{scope}

\begin{scope}[shift={(-2,-6)}]
	\draw[scalarnoarrow](-2.5,1) -- (-1,0);
	\draw[scalarnoarrow](-2.5,-1) -- (-1,0);
	\draw[scalarnoarrow](-1,0) -- (1,0);
	\draw[line](1,0) -- (2.5,1);
	\draw[line](1,0) -- (2.5,-1);
    \node at (-3,1.0) {$\chi$};
	\node at (-3,-1.0) {$\chi$};
    \node at (-0.1,0.46) {$h_i$};
	\node at (1.5,1.2) {$X$};
    \node at (1.5,-1.2) {$X$};
\end{scope}

\begin{scope}[shift={(6,-6)}]
	\draw[scalarnoarrow](-3,1) -- (-1,1);
	\draw[scalarnoarrow](-3,-1) -- (-1,-1);
	\draw[scalarnoarrow](-1,1) -- (-1,-1);
	\draw[scalarnoarrow](-1,1) -- (1,1);
	\draw[scalarnoarrow](-1,-1) -- (1,-1);
    \node at (-3.5,1.0) {$\chi$};
	\node at (-3.5,-1.0) {$\chi$};
    \node at (-0.7,0) {$\chi$};
	\node at (1.2,1) {$h_i$};
    \node at (1.2,-1) {$h_j$};
\end{scope}

\begin{scope}[shift={(13,-6)}]
	\draw[scalarnoarrow](-3,1) -- (-1,1);
	\draw[scalarnoarrow](-3,-1) -- (-1,-1);
	\draw[scalarnoarrow](-1,1) -- (-1,-1);
	\draw[scalarnoarrow](-1,1) -- (1,-1);
	\draw[scalarnoarrow](-1,-1) -- (1,1);
    \node at (-3.4,1.0) {$\chi$};
	\node at (-3.4,-1.0) {$\chi$};
    \node at (-0.7,0) {$\chi$};
	\node at (1.3,1) {$h_i$};
    \node at (1.3,-1) {$h_j$};
\end{scope}

\begin{scope}[shift={(19,-6)}]
	\draw[scalarnoarrow](-2.5,1) -- (-1,0);
	\draw[scalarnoarrow](-2.5,-1) -- (-1,0);
	\draw[scalarnoarrow](-1,0) -- (0.5,1);
	\draw[scalarnoarrow](-1,0) -- (0.5,-1);
    \node at (-3,1.0) {$\chi$};
	\node at (-3,-1.0) {$\chi$};
	\node at (1,1) {$h_i$};
    \node at (1,-1) {$h_j$};
\end{scope}
\end{tikzpicture}
\caption{ Tree level diagrams relevant for the calculation of the relic abundance of $\psi$ and $\chi$. Here $X$ refers to an SM particle, including $h_i$ with $i= 1,2$.} \label{diagrams}
\end{figure}

\section{Relic abundances and cross sections}\label{relicabundance}
In the following, we analyze the relic abundance of the two-component DM scenario previously described obtained via freeze-out of each DM species, analyzing the hierarchy between the DM particles in order to distinguish the bouncing effect. Furthermore, we analyze cross-sections at low temperature relevant for indirect detection observables.

\subsection{Boltzmann equations}
We assume that in the early Universe both DM candidates were in thermal equilibrium with the SM particles. In Fig.~\ref{diagrams} we show the corresponding Feynman diagrams that participate in the relic density calculation. Based on \cite{Belanger:2014vza} and without loss of generality assuming that $m_\psi < m_\chi$, the evolution of the individual singlet abundances $Y_i \equiv n_i/s$, with $i=\psi , \chi$, as a function of the temperature $x \equiv \mu /T$, with $\mu = m_\psi m_\chi /(m_\psi + m_\chi)$, are given by 
\begin{eqnarray}\label{boltza}
 \frac{dY_\psi}{dx} &=& - \lambda_{\psi\bar{\psi}XX}\left(Y_\psi^2 - Y_{\psi ,e}^2\right) + \lambda_{\chi\chi\psi\bar{\psi}}\left(Y_\chi^2 - Y_\psi^2\frac{Y_{\chi ,e}^2}{Y_{\psi,e}^2}\right) - \sum_{i=1,2}\lambda_{\psi\bar{\psi}\chi h_i}\left(Y_\psi^2 - Y_\chi\frac{Y_{\psi ,e}^2}{Y_{\chi,e}}\right),  \\ \label{boltzb}
 \frac{dY_\chi}{dx} &=& - \lambda_{\chi\chi XX} \left(Y_\chi^2 - Y_{\chi,e}^2\right) - \lambda_{\chi\chi\psi\bar{\psi}}\left(Y_\chi^2 - Y_\psi^2\frac{Y_{\chi ,e}^2}{Y_{\psi,e}^2}\right) \nonumber \\ 
                      &+& \frac{1}{2}\sum_{i=1,2}\left[\lambda_{\psi\bar{\psi}\chi h_i}\left(Y_\psi^2 - Y_\chi\frac{Y_{\psi ,e}^2}{Y_{\chi,e}}\right)  -  \lambda_{\chi\psi\psi h_i}\left(Y_\psi Y_\chi  - Y_\psi Y_{\chi,e}\right)\right],
\end{eqnarray}
where we have defined
\begin{align}
	\lambda_{abcd}(x):=
	\dfrac{\langle\sigma_{abcd}v\rangle(x)\cdot s(T)}
	{x\cdot H(T)}, \qquad \text{for} ~ a,b,c,d=\psi,s,h_1,h_2,X,
	\label{def_lambda}
	\end{align}
where $X$ here represents an SM particle. The entropy density $s$ and Hubble rate $H$ in a radiation-dominated universe are given by
	\begin{align}\label{Hubbleandentropy}
	H(T)=\sqrt{\dfrac{4\pi^3G}{45}g_{*}(T)} \cdot T^2, \quad 
	s(T)=\dfrac{2\pi^2}{45}g_{*s}(T)\cdot T^3,
	\end{align}
where $G$ is the Newton gravitational constant, and $g_*$ and $g_{s*}$ are the effective degrees of freedom contributing respectively to the energy and the entropy density at temperature $T$, respectively. The equilibrium densities, $Y_{i,e} \equiv n_{ie}/s$, are calculated using the Maxwell-Boltzmann distribution, whose number density is given by:
	\begin{align}\label{eqden}
	n_{i,e}(T) = g_i\dfrac{m_i^2}{2\pi^2}TK_2(\tfrac{m_i}{T}),
	\end{align}
with $g_i$ the internal spin degrees of freedom, and $K_2$ is the modified Bessel function of the second kind. The equation gets modifications as the hierarchy of the singlets changes, and in this way, the equation may be derived using the detailed balance principle $n_{a,e}n_{b,e}\braket{\sigma_{abcd} v} = n_{c,e}n_{d,e}\braket{\sigma_{cdab} v}$, with the indexes $a,b,c,d$ for the respective particles. We implemented the model into \texttt{LanHEP} \cite{Semenov:2002jw} and into \texttt{micrOMEGAs} code \cite{Belanger:2013ywg} to perform the calculations. For the rest of the paper, sometimes it will be convenient to use the quantity $\Delta_i = 2m_\psi - m_\chi - m_{h_i}$. 

\subsection{Mass hierarchies}\label{masshier}
\begin{figure}[t!]
\centering
\includegraphics[width=0.32\textwidth]{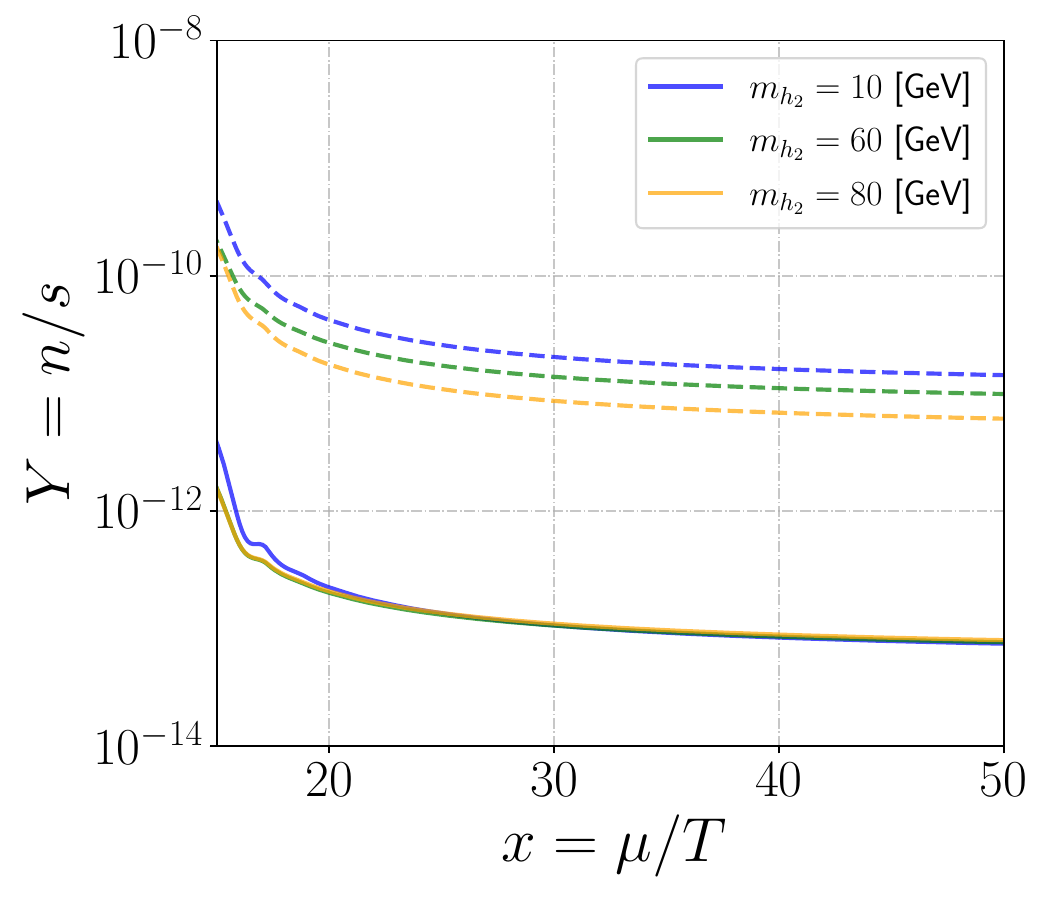}
\includegraphics[width=0.32\textwidth]{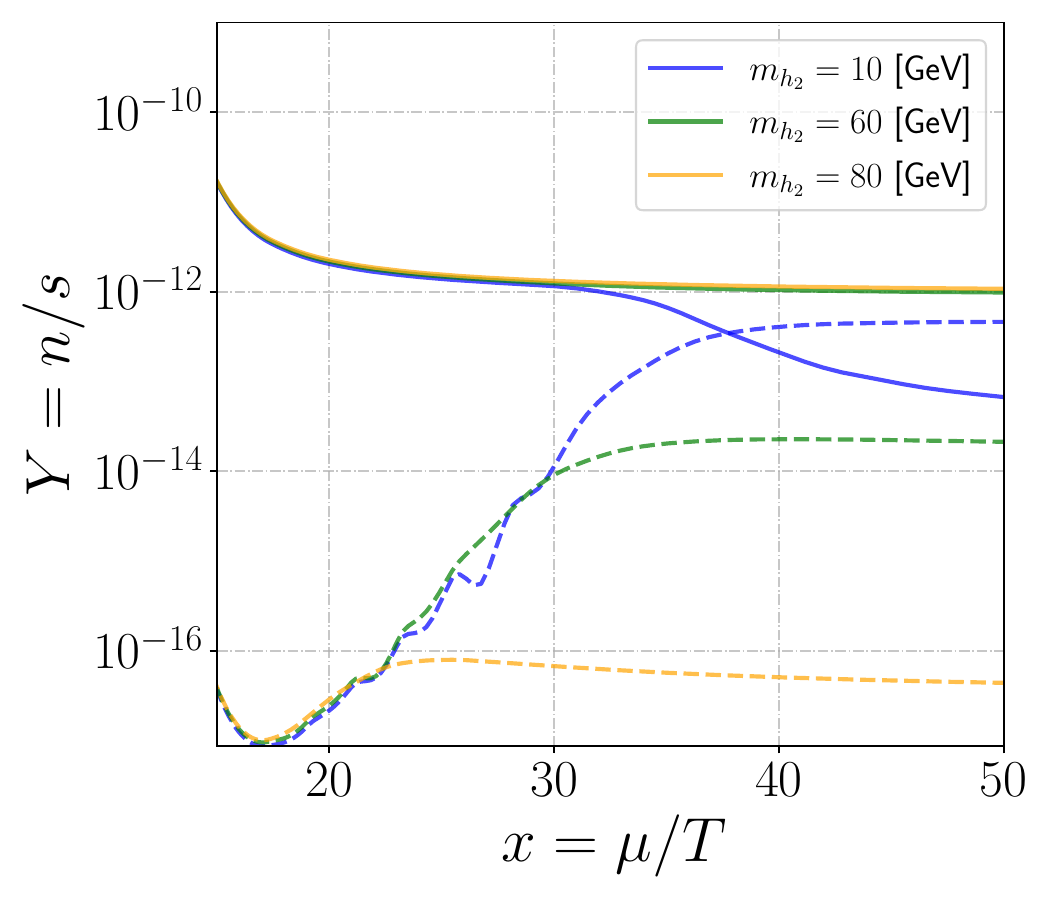}
\includegraphics[width=0.32\textwidth]{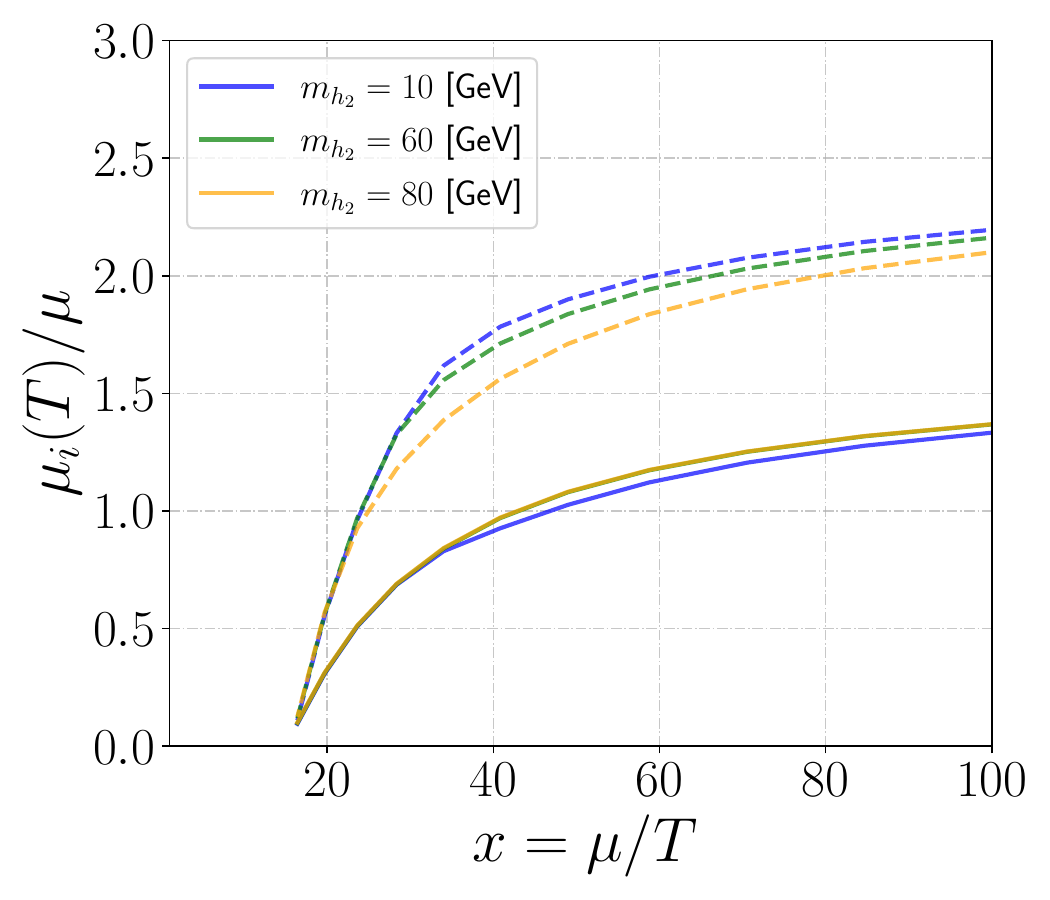}
\caption{In the first two plots we show the yields of $\psi$ (solid lines) and $\chi$ (dashed lines), assuming $g_\psi = 1$ and $\theta = 0.1$. In the plot in the left we consider the normal hierarchy with $(m_\psi, m_\chi) = (200, 150)$ GeV, whereas in the plot in the middle, the inverse hierarchy with $(m_\psi, m_\chi) = (100, 150)$ GeV. The plot on the right shows the evolution of the chemical potentials of each DM particle.}
\label{yields1}
\end{figure}

We distinguish two hierarchies between the DM particles which can make the yield behavior quite different, particularly for $\chi$: (i) $m_\psi > m_\chi$, what we call the \textit{normal hierarchy}, and (ii) $m_\psi < m_\chi$, the \textit{inverse hierarchy}. In the former case, the freeze-out of the DM particles results to follow the standard freeze-out of two interacting DM particles, each one decoupling from the SM plasma at $x \approx 15 - 20$. We have exemplified this with a few parameter space points in Fig.~\ref{yields1}(left) where we show the yields of each DM candidate as a function of the temperature. As the two DM particles present several interactions, they continue decreasing their yields for some time after each particle completely freezes out. 

On the other hand, in the inverse hierarchy, and at high temperatures, $\psi$ and $\chi$ being in thermal equilibrium with the SM plasma, we assume that both particles have vanishing chemical potentials, $\mu_\psi = \mu_\chi = 0$\footnote{In other words, before chemical departure, there is no primordial asymmetry in the fermion sector (see for instance \cite{Graesser:2011wi}). Furthermore, since universally the Greek letter ``$\mu$" is used for both reduced mass and chemical potential, we distinguish each case depending on whether it has a sub-index or not. That is, in the text, $\mu$ will always be reduced mass, whereas $\mu_X$, with $X$ being any particle, will refer to a chemical potential.}. Since $m_\chi > m_\psi$, the fermion is less Boltzmann suppressed than $\chi$, and assuming that $\psi$ does not rise in abundance and they keep the same chemical potential, one has that $n_\chi = (n_{\chi,e}/n_{\psi, e})n_\psi \sim e^{-(m_\chi - m_\psi)/T} n_\psi$, i.e., the abundance of $\chi$ decreases as $T$ decreases. After chemical decoupling from the SM sector, the DM particles may develop a chemical potential $n_i \approx n_{i,e} e^{\mu_i/T}$, in such a way that the yield of $\chi$ now results in
\begin{eqnarray}\label{chi_yield}
 n_\chi \approx \frac{n_{\chi, e}}{n_{\psi ,e}}e^{(\mu_\chi - \mu_\psi)/T} n_\psi .
\end{eqnarray}
Provided $\mu_\chi > \mu_\psi$ and even having a decreasing $n_\psi$, the Boltzmann suppression in eq.~\ref{chi_yield} from the factor $n_{\chi,e}/n_{\psi,e}$ can be compensated by the exponential $e^{(\mu_\chi - \mu_\psi)/T}$, rising exponentially the number density of $\chi$ for some time until Hubble expansion dominates. After the dark particles decouple chemically from the SM thermal bath, the process sustaining chemical equilibrium within the dark sector are semi-annihilations $\psi\bar{\psi}\leftrightarrow \chi h_i$, with $i=1,2$, such that $\mu_\chi \approx 2\mu_\psi$. The resulting effect is shown in Fig.~\ref{yields1} (middle) for a specific choice of parameters, where the rising of $Y_\chi$ results for a finite interval of temperature. The size of the yield increment is model-dependent, and as it can be noted in Fig.~\ref{yields1} (middle), the height of the bouncing depends on $m_{h_2}$. The decreasing in the yield of the bouncing comes from the kinematical suppression of the process responsible for it, i.e. $\psi\bar{\psi}\rightarrow \chi h_2$, since $2m_\psi < m_\chi + m_{h_2}$, suppressing $\braket{\sigma_{\psi\psi\chi h_2} v}$, in turn decoupling $\chi$ and $\psi$ at earlier times in comparison to the case with lower $m_{h_2}$. We say that for those cases we have $\Delta
_2 < 0$ (for this particular case, we also have $\Delta_1 < 0$)\footnote{Notice that, after the bouncing of $\chi$ shown in each case of Fig.~\ref{yields1}(middle), the orange case takes a slightly different behavior. In that case, $Y_\chi$ continues decreasing as the temperature decreases, due exclusively to semi-annihilations of the type $\psi\chi \rightarrow \psi h_i$, this is, the last term in eq.~\ref{boltzb}.}. In Fig.~\ref{yields1} (right), we observe graphically the non-zero values taken by the chemical potential of the two stable particles, fulfilling $\mu_\chi \approx 2\mu_\psi$, with the highest value of $\mu_\chi - \mu_\psi$ being the case with the strongest bouncing. To obtain each chemical potential, we have solved $\mu_i$ from $n_i \approx n_{i,e} e^{\mu_i/T}$, such that 
\begin{eqnarray}
    \mu_i(T) \approx T \log\left(\frac{Y_i}{Y_{ie}}\right), \quad i=\psi,\chi
\end{eqnarray}
with $Y_i(T)$ obtained with \texttt{micrOMEGAs} code after solving the coupled Boltzmann equations (cBE), eq.~\ref{boltza}.

\begin{figure}[t!]
\centering
\includegraphics[width=0.32\textwidth]{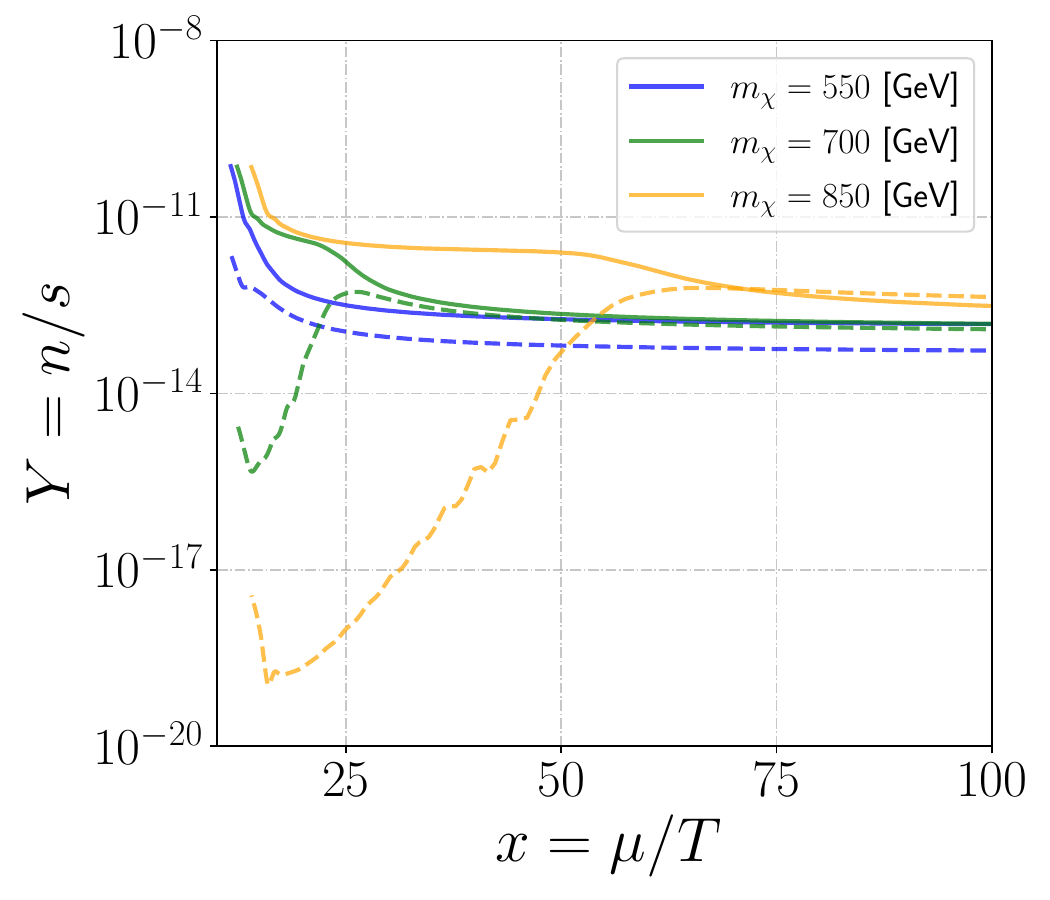}
\includegraphics[width=0.32\textwidth]{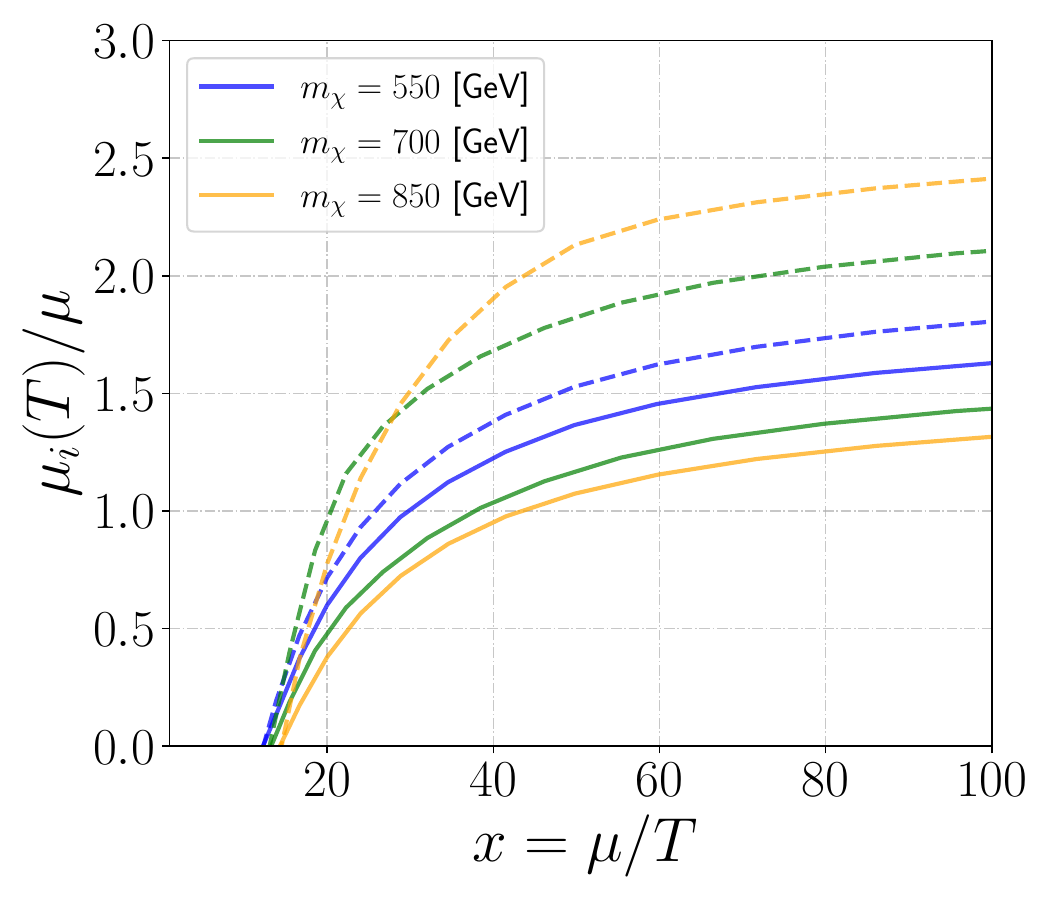}
\includegraphics[width=0.32\textwidth]{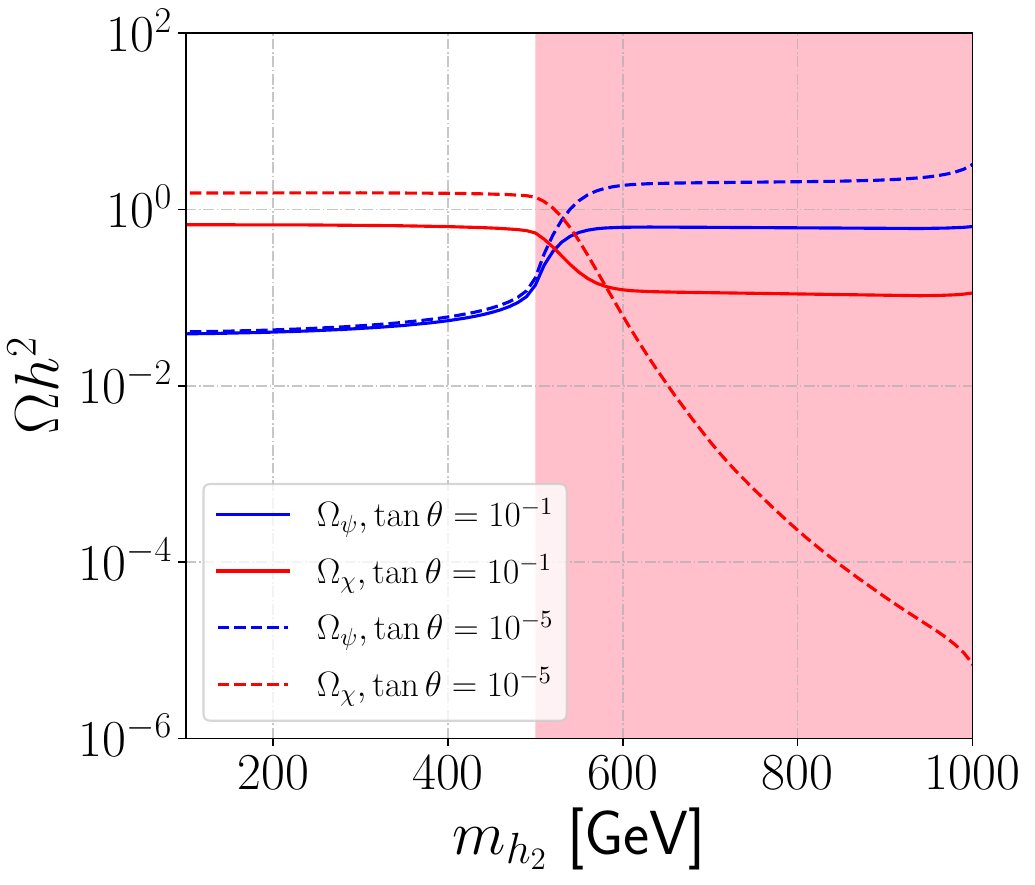}
\caption{(left) Yield behavior as a function of temperature in the inverse hierarchy. Here we set $m_\psi = 500$ GeV, $m_{h_2} = 600$ GeV, $g_\psi = 1$ and $\tan\theta = 0.1$. (middle) Chemical potential for each stable particle considering the same parameters as in the left plot. (right) Relic abundance for $\psi$ and $\chi$, assuming $m_\psi = 1000$ GeV, $m_\chi = 1500$ GeV, $g_\psi = 1$. The pink region represents $\Delta_2 < 0$.}
\label{yields2}
\end{figure}

Another relevant feature regarding the bouncing is the variation in the minimum $Y_\chi$ at which the bouncing starts to appear. If the pNGB remains more time in thermal equilibrium with the SM, the bouncing will start at later times, and it may grow the yield several orders of magnitude before coming to an end. This is exemplified in Fig.~\ref{yields2}(left) for the parameters indicated in the plot. As shown, in this case, it is clear that the bigger $m_\chi$, the bigger the bouncing. This effect is due to the fact that the semi-annihilation term is proportional to  $\lambda_{\psi\bar{\psi}\chi h_i}$ and impacts directly in the decoupling temperature of $\chi$ from the thermal bath, with a strong dependence on $m_\chi$. In App.~\ref{AppendixA} we study this effect in greater detail from a numerical and semi-analytical way.

As explained before, the bouncing effect proceeds through either semi-annihilation $\psi\bar{\psi}\rightarrow \chi h_1$ or $\psi\bar{\psi}\rightarrow \chi h_2$, each process with a dependence of $\sin^2\theta$ and $\cos^2\theta$, respectively, affecting the relic abundance directly. For instance, in Fig.~\ref{yields2}(right) we plot the relic abundance as a function of the second Higgs mass, for $m_\psi = 1000$ GeV, $m_\chi = 1500$ GeV and $g_\psi = 1$, for different values of $\theta$. For this parameter choice, the bouncing effect is present for all the shown combinations of values of $(m_{h_2}, \theta)$. As it is shown by the solid red line ($\theta = 0.1$), $\Omega_\chi$ remains constant for $m_{h_2} \gtrsim 500$ GeV since, even though $\braket{\sigma v}_{\psi\bar{\psi}\chi h_2}$ is kinematically suppressed (i.e., the pink region representing $\Delta_2 < 0$), the process  $\psi\bar{\psi}\rightarrow \chi h_1$ becomes the leading one for the bouncing effect. On the contrary, for a more decoupled dark sector-SM case, e.g. $\theta = 10^{-5}$, $\psi\bar{\psi}\rightarrow \chi h_1$ process becomes suppressed by $\sin\theta$, and $\psi\bar{\psi}\rightarrow \chi h_2$ remains as the effective process, with $\Omega_\chi$ decreasing as $m_{h_2}$ gets higher values\footnote{We have checked that for such small $\theta$ values the dark sector remains still in thermal equilibrium with the SM  through the comparison of particle reaction rates with the Hubble expansion rate in a radiation dominated universe before the onset of chemical decoupling of the dark relics. For thermalization effects in Higgs portal with singlet-doublet Higgs mixing see also \cite{Krnjaic:2015mbs, Yang:2022zlh, Amiri:2022cbv}, where similar conclusions were obtained, i.e. the thermalization is lost in the ballpark of $\theta \sim 10^{-7} - 10^{-6}$, for GeV scale mass of $h_2$.}. In that case, the bouncing effect becomes smaller. Therefore, the mixing angle and $m_{h_2}$ are important parameters behind the bouncing effect, impacting the value of $\Omega_\chi$, and also determining the leading semi-annihilation in the relic density calculation.
\begin{figure}[t!]
\includegraphics[width=1\textwidth]{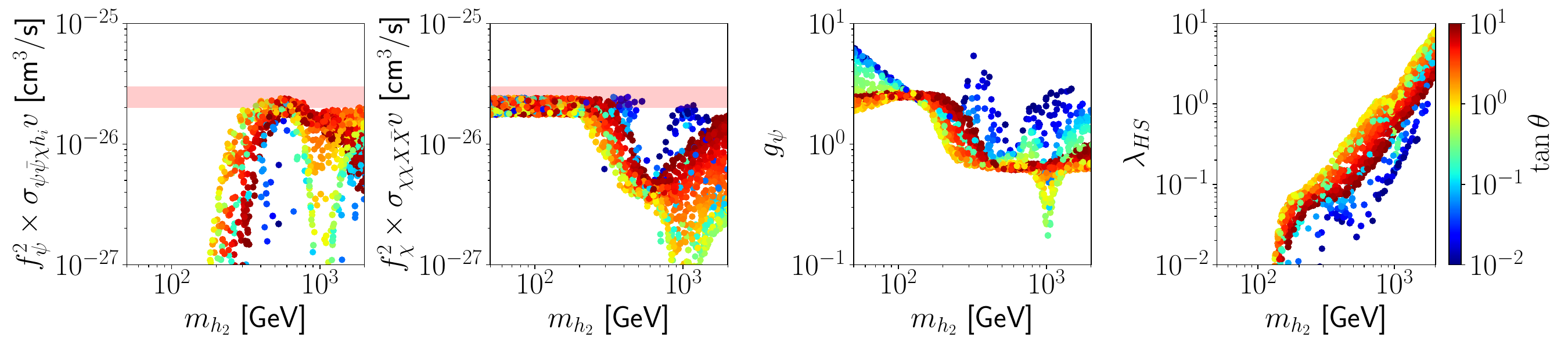}
\includegraphics[width=1\textwidth]{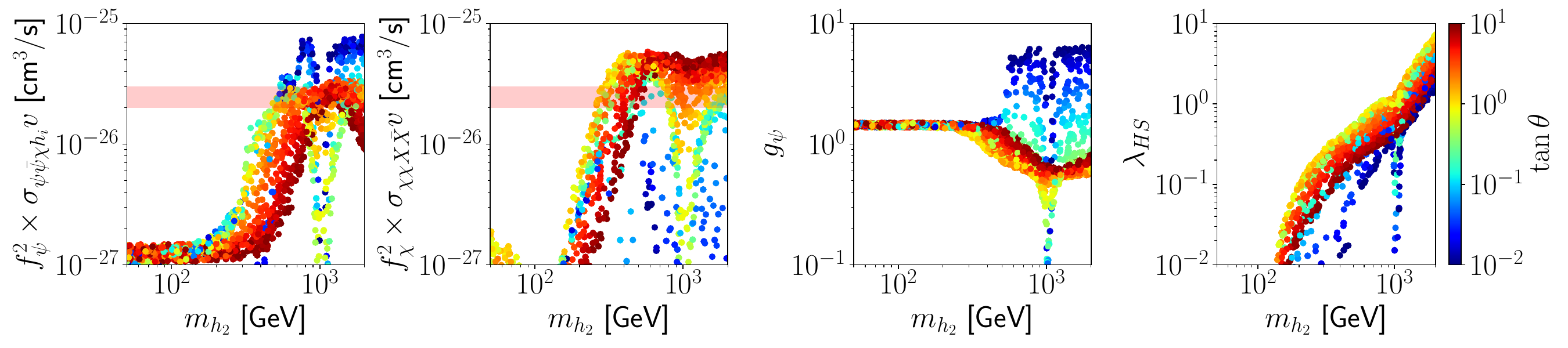}
\caption{Random scans with each point fulfilling the correct relic abundance, assuming $m_\psi = 500$ GeV, and the color indicating the value of $\theta$. In the first row we consider $m_\chi = 300$ GeV, whereas in the second one $m_\chi = 700$ GeV. In the first two $y$-axes, $f_\text{i} \equiv \Omega_\text{i}/\Omega_c$, with $\Omega_\text{i}$ the relative abundance of the $i = \psi, \chi$ initial state. For more details of the scan see the text.}
\label{random}
\end{figure}
\subsection{Cross sections}\label{crosssections}
In this subsection, we study the values of some of the most relevant average annihilation cross section times relative velocity at $v_\text{rel}\rightarrow 0$ from DM (semi)annihilation, especially when they surpass the thermal canonical value $2-3\times 10^{-26}$ cm$^3/$s \cite{Puetter:2022ucx}. We run two random scans, one considering $m_\chi = 300$ GeV and the other $m_\chi = 700$ GeV. In both scans we have considered $m_\psi = 500$ GeV, $m_{h_2} = [50, 2000]$ GeV, $g_\psi = [0.1, 10]$ and $\tan\theta = [10^{-2}, 10^1]$. We have selected all the points which match the observed relic abundance. As shown in the first row of Fig.~\ref{random}, we have projected the points in different planes, with the color of each point indicating the corresponding value of $\tan\theta$. As it is expected in the normal hierarchy $m_\psi > m_\chi$, the first two plots corroborate the fact that the weighted cross sections never surpass the thermal cross section reference (pink regions)\footnote{To obtain the average cross sections, we have used \texttt{micrOMEGAs} making use of the functions vs1120F(T) and vs2200F(T), where $1$ and 2 refers to $\psi$ and $\chi$, respectively. We have taken $T = T_\text{end} = 10^{-3}$ GeV, a temperature small enough to consider the processes in their non-relativistic regime.}. The other two plots in the right of the first line show the corresponding values for the couplings of the model.

In the lower row of Fig.~\ref{random}, we project the random scan for the inverse hierarchy $m_\psi < m_\chi$. Contrary to the previous case, here many points surpass the thermal reference value in the first two plots, although with clear differences in their value acquired by $\theta$. Notice that for the case of fermion DM annihilation, those points above the thermal relic tend to be disfavored by the high values of $g_\psi$, due to perturbativity. For the case of the pNGB DM annihilation, the points with strong cross-sections seem to respect perturbativity for $g_\psi$ and $\lambda_{HS}$. Nonetheless, as we show in the next section, $\theta \gtrsim 0.1$ enter in conflict with both direct detection and collider constraints.

Finally, a few comments to highlight. First, analogously to the 125 GeV Higgs boson $h_1$, we have been assuming that the mediator $h_2$ remains in thermal equilibrium with the SM plasma at any moment, so its chemical potential vanishes. Secondly, a necessary condition for the existence of the bouncing is $m_\chi >m_\psi$, but it is also present in either case $m_\psi > m_{h_2}$ or $m_\psi < m_{h_2}$. Third, $\Delta_1 > 0$ and/or $\Delta_2 > 0$ do not guarantee the bouncing effect. In the following section we study the phenomenology of the model.

\section{Phenomenology}\label{pheno}
In this section we consider the setup at hand as a fully realistic DM model, considering constraints from collider and DM searches. We explore the viability of pNGB DM below 50 GeV, and also we study the normal and inverse hierarchy for masses of hundreds of GeV. Finally, we explore a few indirect detection observables.

\subsection{Constraints}
\subsubsection{Theoretical constraints}
We consider perturbativity on the couplings, i.e. $g_\psi \leq \pi$ and $\lambda_{HS} \leq \pi$.
\subsubsection{Dark matter constraints}
The relic abundance measure today is given by the most updated Planck result \cite{planck2018}. We take the value $\Omega_c h^2 = 0.12$ as the measured DM relic density, with an error of 10\% in the calculation of it with \texttt{micrOMEGAs}. As usual, the total relic abundance in the two-component DM model is given by the sum of each stable particle, i.e. $\Omega h^2 = \Omega_\psi h^2 + \Omega_\chi h^2$.

On the other hand, direct detection (DD) for the pNGB DM candidate is relaxed due to its Goldstone nature \cite{Gross:2017dan} (unless new explicit global symmetry-breaking sources are present \cite{Alanne:2020jwx}). It has been shown that the direct detection rate of a pNGB DM particle is too low to be observed with present experiments \cite{Gross:2017dan}, even at the one-loop level \cite{Ishiwata:2018sdi}, then we do not take into account the relative contribution of the pNGB for DD constraints\footnote{We have checked with \texttt{micrOMEGAs} that the resulting SI cross section for the pNGB DM is always below the bounds given by present DD bound experiments.}. On the contrary, $\psi$ is subject to sizable constraints appearing from the scattering in $t$-channel exchange of $h_1$ and $h_2$. At tree level, it is given by \cite{Garcia-Cely:2013wda}
\begin{eqnarray}\label{dd}
 \sigma_\psi^{SI} \approx \frac{m_N^4m_\psi^2 f_p^2}{4\pi v_h^2(m_\psi + m_N)^2}\left(\frac{1}{m_{h_1}^2} - \frac{1}{m_{h_2}^2}\right)^2(g_\psi \sin 2\theta)^2 ,
\end{eqnarray}
where $m_N$ denotes the nucleon mass and $f_p = f_n \approx 0.27$. Even though in most of the cases we take the most recent bounds on direct detection from Lux-Zeplin experiment (LZ) \cite{LZ:2022ufs}, in some cases we also used XENON1T \cite{Aprile_2018} and XENONnT projection \cite{XENON:2020kmp}. As an example of the intensity of LZ constraints on the parameter space of the model, in Fig.~\ref{constraints}(left) we show the resulting allowed parameter space for a representative fermion mass of 500 GeV, taking $g_\psi = \pi$ (black lines) and $g_\psi = 0.5$ (grey lines). The region above each curve is excluded by LZ, and the corresponding dashed lines represent the bound when the fermion acquires 10\% of the total relic abundance. As it can be appreciated, direct detection bounds are strong, particularly for $m_{h_2} < m_{h_1}$.

\subsubsection{Collider constraints}
A second Higgs is constrained throughout the combination of its mass and coupling to the SM particles, i.e. its mixing angle. Direct searches of a second Higgs set that $\theta \lesssim 0.15$ for $m_{h_2} < 50$ GeV, whereas for $m_{h_2} > 100$ GeV, electroweak precision tests (EWPT) set that $\theta \lesssim 0.3$ \cite{Falkowski:2015iwa, Arcadi:2016qoz}. If either of the DM particles has a mass below half of the mass of the 125 GeV Higgs boson, the latter may decay into a pair of DM particles, resulting in an invisible decay of the Higgs boson at colliders. Measurements set limits on the branching fraction of the Higgs into invisible particles, with the most updated value being BR$(h_1\rightarrow \text{invisible}) \lesssim 14\%$ at 95$\%$ C.L. \cite{ATLAS:2022yvh}. In our scenario, if the kinematic is allowed, we have that 
\begin{eqnarray}
 \Gamma(h_1\rightarrow \text{invisible}) = \Gamma(h_1\rightarrow \psi\psi) + \Gamma(h_1\rightarrow \chi\chi)
\end{eqnarray}
where each decay width is given by
\begin{eqnarray}
 \Gamma(h_1\rightarrow \psi\bar{\psi}) &=& \frac{g_\psi^2}{16\pi}\left(1 - \frac{4}{r^2}\right)^{3/2}m_{h_1}\sin^2\theta ,\\
 \Gamma(h_1\rightarrow \chi\chi) &=& \frac{g_\psi^2 r^2}{32\pi}m_{h_1}\sin^2\theta \sqrt{1 - 4\frac{m_\chi^2}{m_{h_1}^2}} , 
\end{eqnarray}
with $r \equiv m_{h_1}/m_\psi$. For the parameter space that we are interested in, $h_2$ is short-lived \footnote{To have $\tau_{h_2} > 1$ s for $h_2$ masses in the $\gtrsim\mathcal{O}(\text{GeV})$, it should occur that $\theta \lesssim 10^{-10}$.}, therefore the decay $h_1 \rightarrow h_2h_2$ gives not invisible products, but it is important for the total width decay of the Higgs boson. Furthermore, as $h_2$ is short-lived, it does not affect constraints from Big Bang Nucleosynthesis (BBN). Finally, electroweak precision test set constraints on the combination $(m_{h_2},\theta)$, but for $m_{h_2} < m_{h_1}/2$ they result to be much weaker than Higgs to invisible constraints \cite{Falkowski:2015iwa}.

\subsection{pNGB mass below $m_h/2$}
In the original scenario of pNGB DM \cite{Gross:2017dan}, masses for the latter below $\sim 50$ GeV are normally not possible, since Higgs to invisible bounds on the Higgs portal coupling become relevant, and to fulfill the correct relic abundance, the Higgs portal can not take arbitrarily small values, otherwise an overabundance is obtained. One way to avoid the latter is having $2m_\chi \sim m_{h_2}$ in such a way that the annihilation of the pNGB occurs on-shell in the $s$-channel, then decreasing sufficiently the Higgs portal coupling to evade the Higgs to invisible constraint (see e.g. \cite{Cline:2019okt}). In our two-component DM scenario, we show that it is possible to have viable pNGB with masses below $m_h/2$ without entering into resonance effects, fulfilling the correct relic abundance, and evading Higgs to invisible bounds and direct detection.
\begin{figure}[t!]
\includegraphics[width=0.32\textwidth]{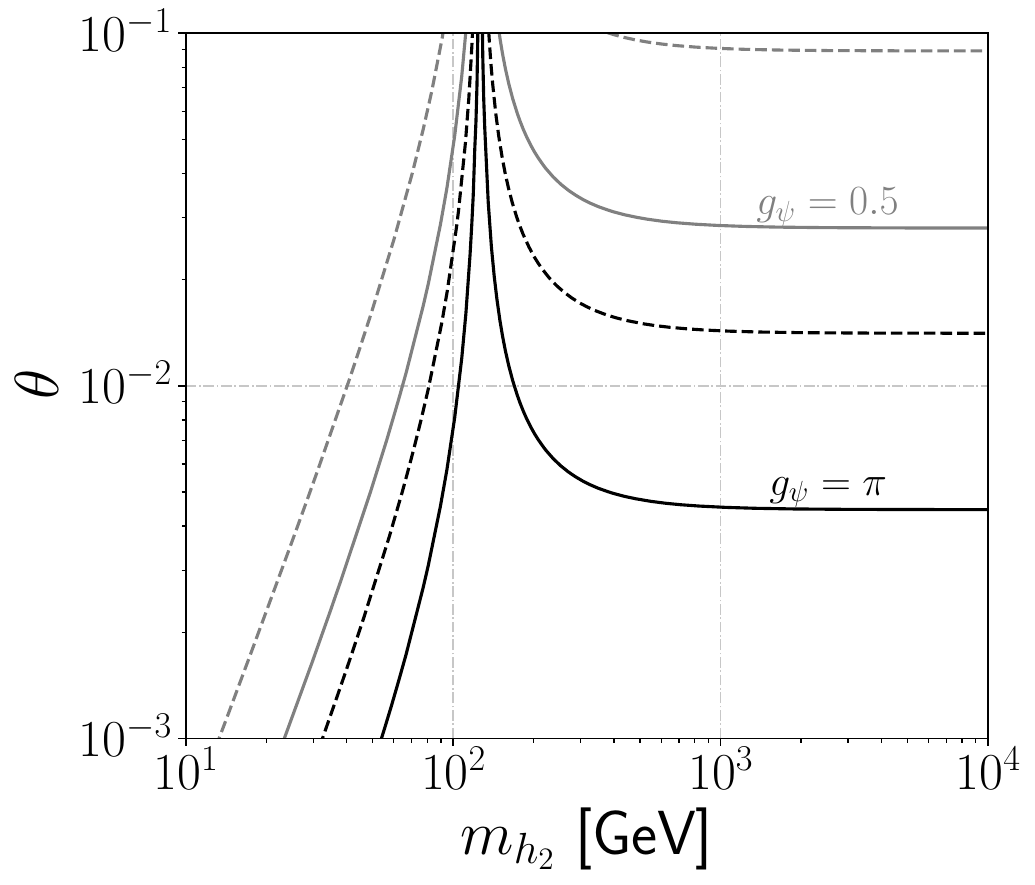}
\includegraphics[width=0.32\textwidth]{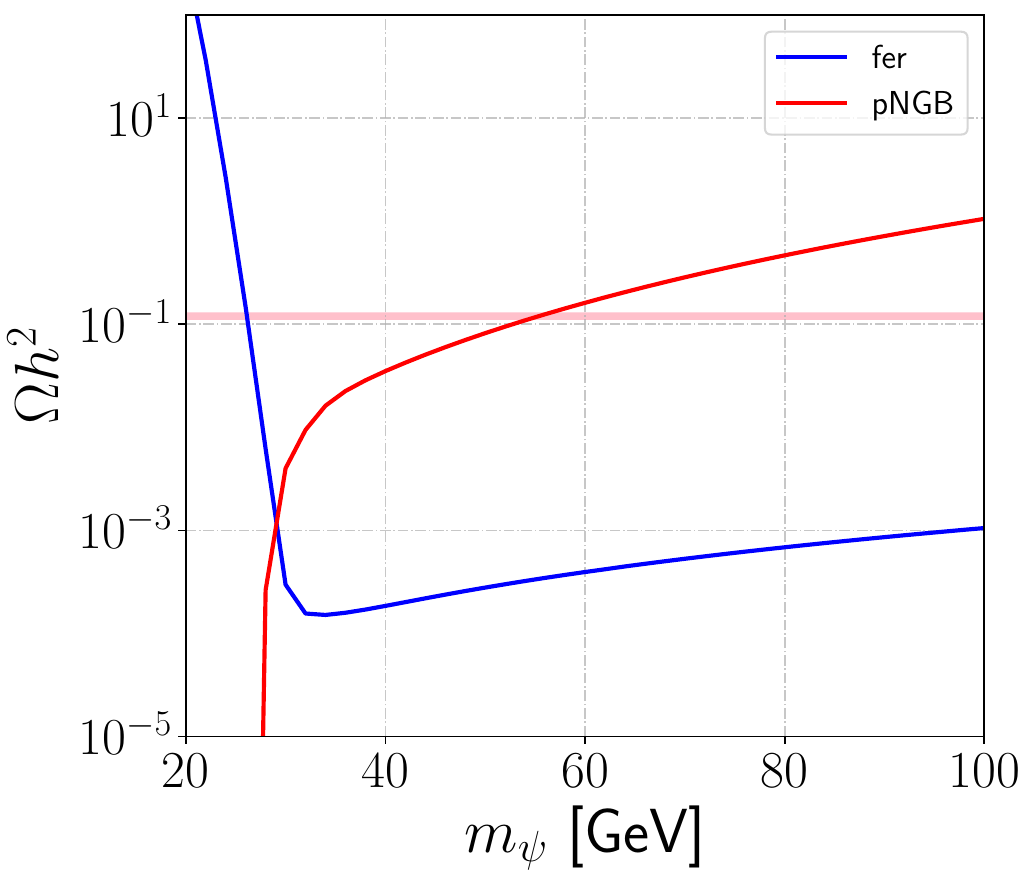}
\includegraphics[width=0.32\textwidth]{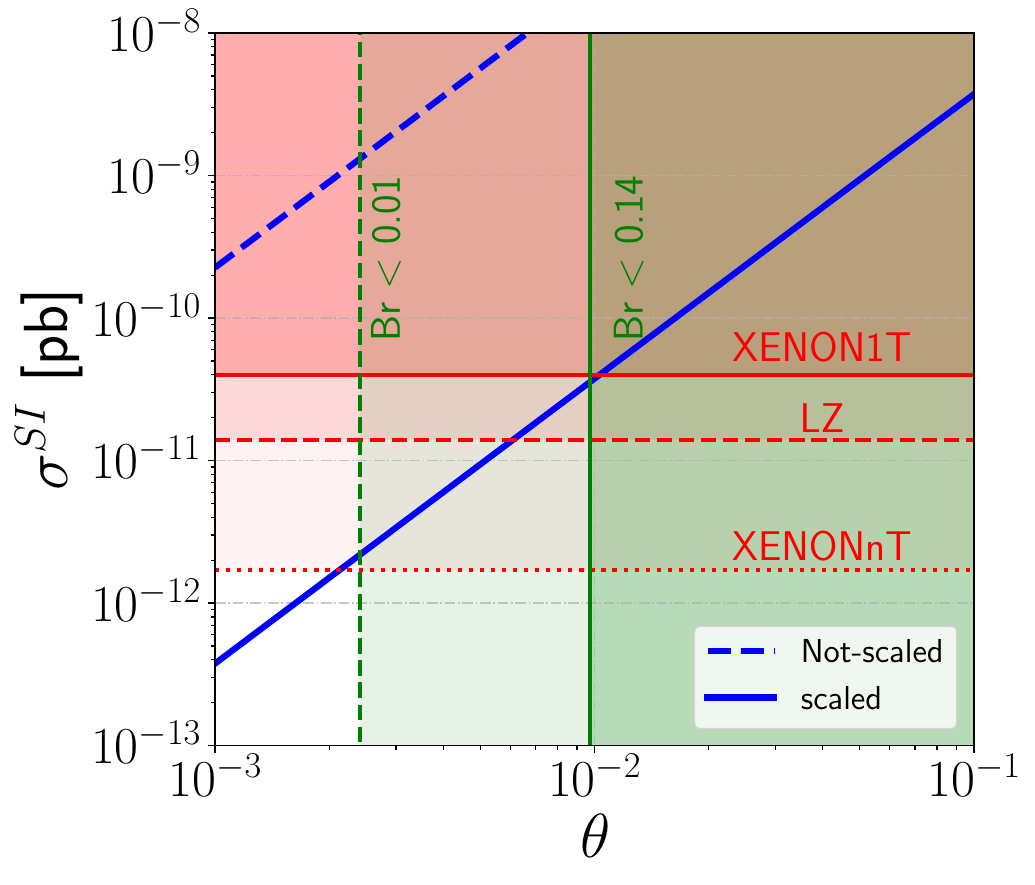}
\caption{(left) Direct detection constraints on the fermion DM considering LZ bounds. The solid black (grey) curves represent the contours of $\sigma^{SI}$ for $m_\psi = $ 500 GeV and $g_\psi = \pi$($0.5$), whereas the respective dashed curves consider that the fermion contributes with a 10\% of the total relic abundance. LZ data rule out the region above each curve. (Middle) Relic abundance in the low-mass regime, assuming $m_\chi = 30$ GeV, $m_{h_2} = 29$ GeV, $g_\psi = 1$, and $\theta < 0.1$. The horizontal pink region corresponds to the observed relic abundance. (right) Direct detection and invisible Higgs decays constraint assuming $m_\psi = 55$ GeV, $m_\chi = 30$ GeV, $m_{h_2} = 29$ GeV and $g_\psi = 1$ (i.e., the parameter space point that in the plot in the middle fulfill the correct relic abundance). In red we show constraints from XENON1T, LZ and the projection of XENONnT, and in green the excluded regions by the invisible Higgs decay considering a branching of 0.14 (dark green) and a projection of 0.01 (light green).}
\label{constraints}
\end{figure}
The dynamics for the calculation of the relic abundance is independent of the mixing angle provided $\theta \lesssim 0.1$, since all the diagrams containing $h_1$ are suppressed by $\tan^2\theta$, then the relic abundance is determined by the fields of the dark sector and $h_2$. In Fig.~\ref{constraints}(middle) we show the relic abundance behavior for the two DM candidates as a function of $m_\psi$, assuming $m_\chi = 30$ GeV, $m_{h_2} = 29$ GeV, $g_\psi = 1$ and $\theta < 0.1$. Here, $m_\psi \approx 55$ GeV is the required value to achieve the correct relic abundance. Notice that $m_{h_2}$ must be lighter than the two DM components, in order to avoid overabundance. In the same line, we observe in the right plot of Fig.~\ref{constraints}, that the maximum value that $\theta$ can take to evade LZ  is $\sim 0.006$. Notice here that the solid line considers the scaling of $\sigma^\text{SI}$ by the fraction of relic abundance of the fermion. It is interesting to observe here that constraints from 125-GeV Higgs decaying into invisible particles are complementary to direct detection (green regions), although less strong than LZ in this case. Notice that future projections from XENONnT and collider searches will be competitive between them, excluding even much more parameter space. 

In summary, it is possible to have pNGB DM for masses below $\sim 50$ GeV, without recurring to resonance effects. The price to pay in order to evade strong direct detection and Higgs to invisible constraints is to decrease $\theta $ enough in such a way to compensate the growing behavior of the spin-independent cross section with light $h_2$, since $\sigma^{SI}\sim \sin^2(2\theta)/m_{h_2}^4$. 

\subsection{Indirect detection prospects}

Considering that the two-DM scenario presents sizable average annihilation cross section at low temperatures, now we focus on specific (semi)annihilations prospects relevant for indirect detection observables. We pay special attention to the (semi)annihilation processes $\psi\bar\psi\rightarrow \chi h_i$ and $\chi\chi\rightarrow XX$, with $X$ an SM state\footnote{For a related study of ID signals of the former cross section, see \cite{Garcia-Cely:2013wda, DiazSaez:2021pmg, Bonilla:2023egs}}. The corresponding partial average annihilation cross-sections have been calculated using \texttt{CalcHEP}, expanding around $v_\text{rel} = 0$, in order to retain the $s$-wave contribution only (see App.~\ref{Appendixcross} for the resulting analytical expressions). 

\begin{figure}[t!]
\centering
\includegraphics[width=1\textwidth]{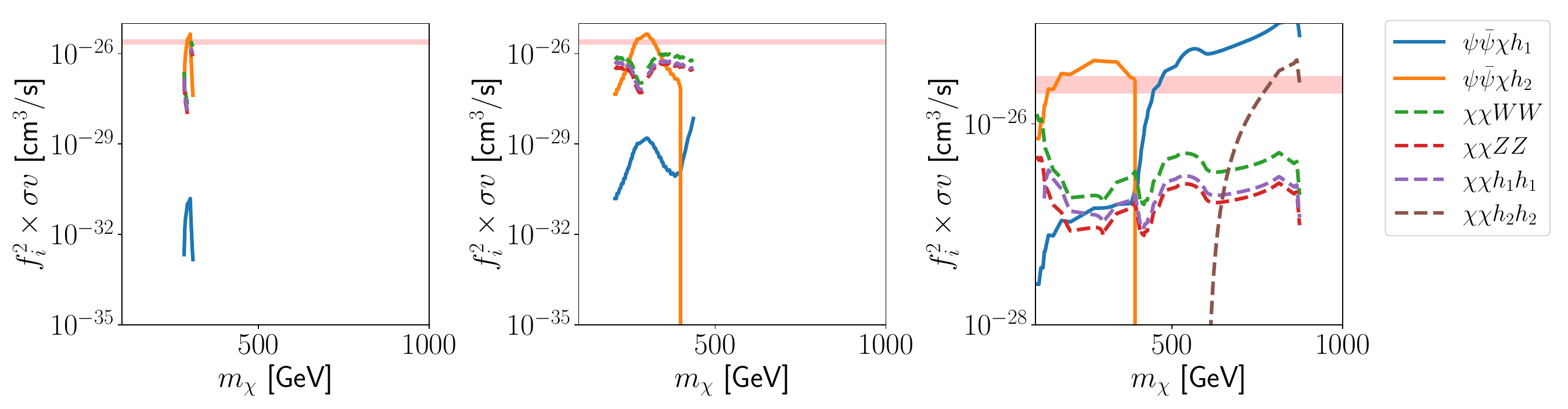}
\includegraphics[width=1\textwidth]{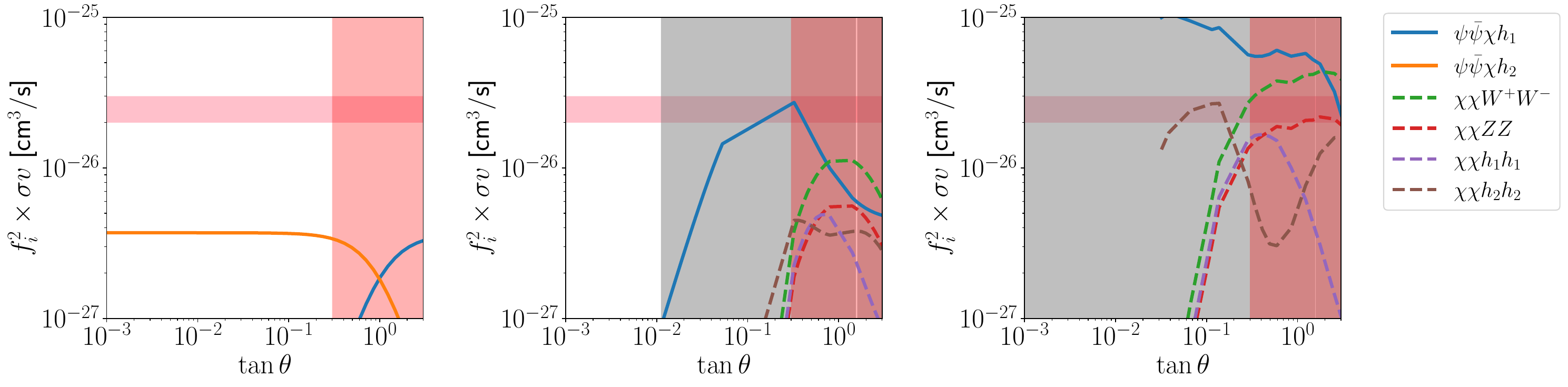}
\caption{(top) Relic weighted average cross section times relative velocity as a function of $m_\chi$. In each plot we have taken $m_\psi = 500$ GeV, $m_{h_2} = 600$ GeV, and $\tan\theta = (10^{-3}, 10^{-2}, 10^{-1})$ (from left to right). $g_\psi$ takes the necessary value to obtain the correct relic abundance, considering perturbativity constraints. The pink band represents the thermal canonical value $2-3\times 10^{-26}$ cm$^3/$s. The solid lines are the values corresponding to the fermion DM annihilation, whereas the dashed lines are the corresponding pNGB annihilation. (bottom) Zero-velocity average annihilation cross section for different DM (semi)annihilation channels as a function of the mixing angle. The values of the parameters here have been taken as $m_\psi = 500$ GeV, $m_\chi = 800$ GeV, $m_{h_2} = (130, 300, 600)$ GeV (from left to right in the plots), and $g_\psi$ takes the necessary value to obtain the correct relic abundance. The pink band is a reference for the thermal value, the grey one represents LZ bounds, and the red one collider constraints.}
\label{id1}
\end{figure}

The partial cross sections result to be highly dependent on the parameters of the model. As we exemplify in the upper row of Fig.~\ref{id1}, considering $m_\psi = 500$ GeV, $m_{h_2} = 600$ GeV and $g_\psi$ getting the appropriate value to match the correct relic abundance, the cross sections not only vary by orders of magnitude depending on $m_\chi$, but as $\theta$ decreases, the parameter space available to obtain the correct relic abundance shrinks allowing only $m_\chi \approx m_{h_2}/2$, otherwise an overabundance is obtained. In this way, in the normal hierarchy and small mixing angles it is possible to obtain sizable cross sections, as shown by the orange solid line and the dashed curves in the left plot of the upper row of Fig.~\ref{id1}, but only in a reduced parameter space. On the contrary, higher $\tan\theta$ values, e.g. $\tan\theta=0.1$, imply less suppression for (semi)annihilation processes into SM states including $h_1$ in the final state, presenting strong cross sections specially in the case $m_\chi > m_\psi$, where the bouncing effect is present. This can be seen in the third plot of the first row of Fig.~\ref{id1}, showing a wider range of $m_\chi$ allowed. For completeness, we also present the case with $\tan\theta = 10^{-2}$. 

We confront the resulting zero-velocity relic weighted cross sections with direct detection bounds from LZ for three scenarios in which we vary $m_{h_2}$, with each case fulfilling the correct relic abundance. In Fig.~\ref{id1}(below), we show the results as a function of the singlet-doublet mixing angle assuming $m_\psi = 500$ GeV, $m_\chi = 800$ GeV, and $m_{h_2} = (130, 300, 600)$ GeV (from left to right, respectively). We have taken small values for $m_{h_2}$ to see the relaxing effect on direct detection bounds. LZ data rule out the shaded region in each plot. In the left plot of Fig.~\ref{id1}(below), LZ bounds are relaxed, due to the algebraic cancellation between the close-in mass of $h_1$ and $h_2$. As $m_{h_2}$ deviates away from $m_{h_1}$, LZ bounds start to be notorious and strong, as it is shown by the middle and the plot in the right, even for $\theta \ll 1$. In this way, all the cross sections with values above the thermal value obtained in the case in which $m_\psi < m_{h_2} < m_\chi$, result disfavored by LZ. Notice also that if we take $m_{h_2} \ll m_{h_1}$, direct detection becomes even stronger, therefore we do not look into that region. In any case, even though direct detection constrains the average cross-section above the thermal value (strongest bouncing effect regime), the model remains viable, with indirect detection near the ballpark of the thermal reference value. A detailed analysis of the full viable parameter space is beyond our goal in this work.

\section{Conclusions}\label{conclusions}
We have studied a simple extension to the SM which under reasonable assumptions presents two DM candidates simultaneously, a fermion and a pNGB, with each one communicating to the SM via the Higgs portal through two Higgs-like bosons. Assuming a thermal scenario in which both DM candidates freeze-out in a radiation-dominated universe, we have found a peculiar yield behavior for the pNGB when this is heavier than the fermion singlet: it \textit{bounces}. This exponential yield increment may reach several orders of magnitude. This model is one of
the first scenarios in which the bouncing effect is exemplified in more
detail.

This effect has a direct impact on indirect searches. In this way, we have explored the zero velocity average annihilation cross sections relevant for indirect searches, finding parameter space regions in which both the fermion semi-annihilation and the pNGB annihilation today present values above the canonical thermal value. We have seen that as the fermion is subject to strong tree-level spin-independent DD constraints, the model requires a suppression on the mixing angle $\theta$ unless the two Higgses become too degenerated. In either case, collider, and specially LZ upper bounds, force to reduce some of the parameters of the model in such a way that the strongest indirect detection signals must be suppressed when they fulfill the correct thermal value.

Last but not least, we have shown that it is possible to recover pNGB DM for masses below 50 GeV, in contrast to the simple model in which invisible Higgs decays ruled out the parameter space for low masses (unless a resonance effect is present). The cost of this is to decrease the singlet-doublet mixing angle to values much below the unity. All in all, the model, being a multi-component DM scenario, is not only elegant in its construction, but it presents the interesting effect of pNGB yield bouncing in the early universe, although the model itself suffers severe constraints from direct detection experiments.

\section{Acknowledgments}
We want to thank to Alfonso Zerwekh, Sebastian Norero, Sasha Belyaev, Claudio Dib, Nicolás Bernal, Jeremy Echeverría, Oscar Zapata, Diego Aristizabal, and Iason Baldes for helpful and encouraging discussions. B.D.S has been founded by ANID (ex CONICYT) Grant No. 3220566. P.E has been founded by PIIC 2022-I, DPP, UTFSM, ANID-Chile Grant 21210952, and ANID-Chile FONDECYT grant No. 1230110. B.D.S also wants to thanks DESY and the Cluster of Excellence Quantum Universe, Hamburg, Germany.
\appendix
\section{Boltzmann equations for the bouncing effect}\label{AppendixA}
\begin{figure}[t!]
\centering
\includegraphics[width=0.3\textwidth]{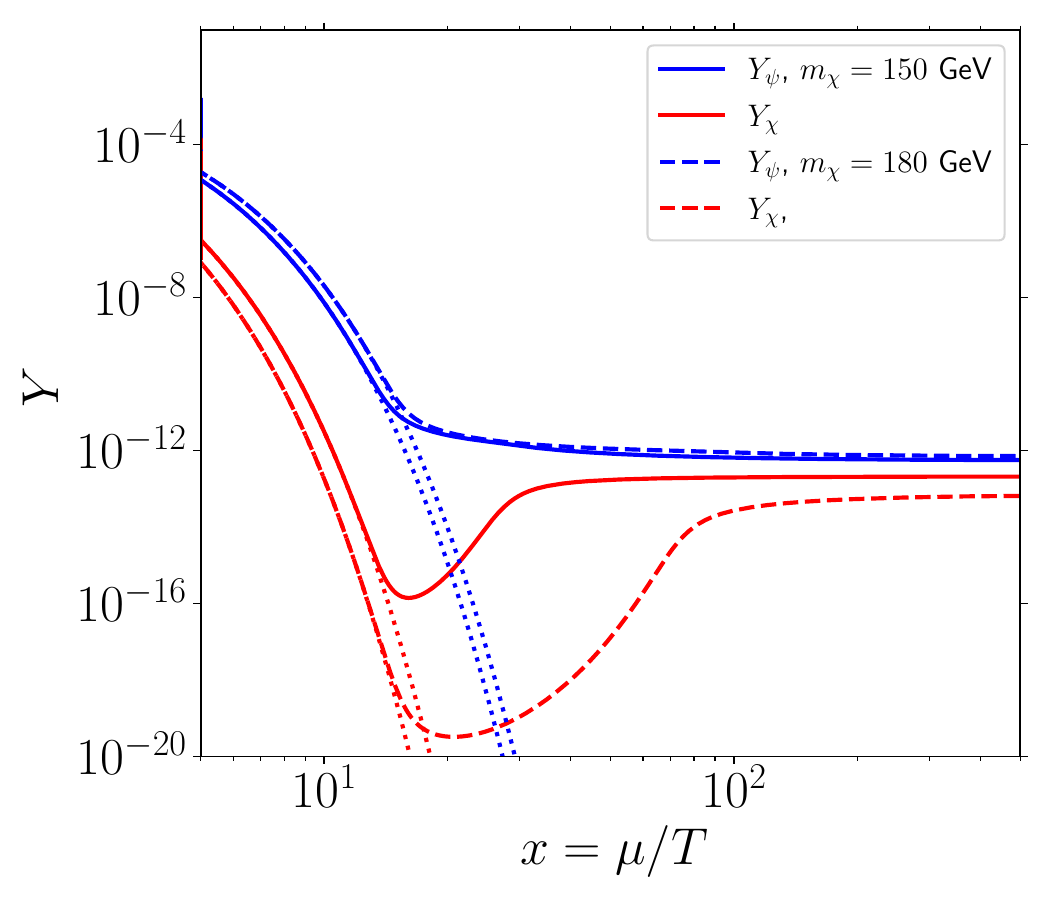}
\includegraphics[width=0.3\textwidth]{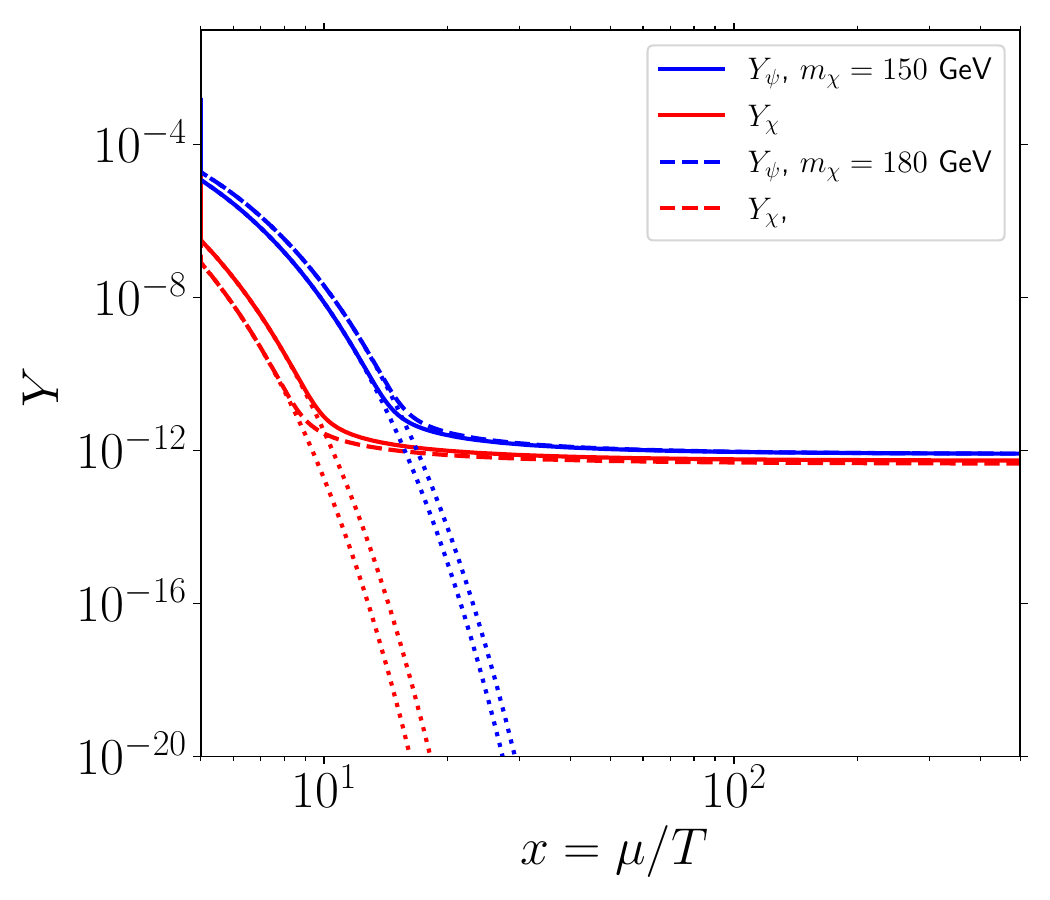}
\includegraphics[width=0.37\textwidth]{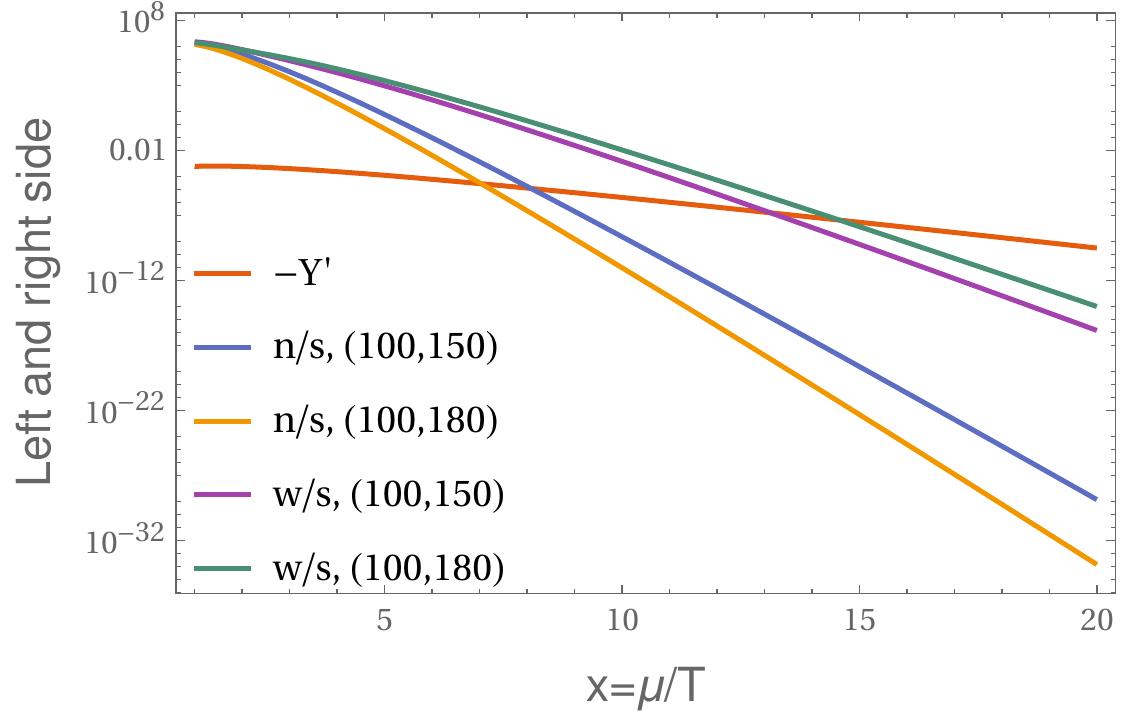} 
\caption{(left and middle) Numerical yield behavior for the cBE in eq.~\ref{cBEappa} and \ref{cBEappb}, assuming $m_\psi = 100$ GeV, $\lambda_{A1} = \lambda_{A2} = \lambda_{S}$, with each cross section set to the canonical value $\braket{\sigma v} = 10^{-9} $ GeV$^{-2}$. The plot in the left considers the semi-annihilation term proportional to $\lambda_S$, whereas the plot in middle does not. (right plot) Solution to the transcendental equation eq.~\ref{foapp}, with the intersection of the orange curve with the rest at $x_f$. In the legend, each pair of values in the parenthesis represents $(m_\psi, m_\chi)$ GeV.}\label{figapp}
\end{figure}

In this appendix we study the interplay and impact of some terms of the cBE in eq.~\ref{boltza} and \ref{boltzb} on the resulting yield evolution. We take a simple approach considering only the first and third terms of each equation, solving them numerically and in a semi-analytical way. As we are interested in the bouncing effect details, we always consider $m_\chi > m_\psi$. For the sake of simplicity, we define $\lambda_{A1} \equiv \lambda_{\psi\psi XX}$, $\lambda_{A2} \equiv \lambda_{\chi\chi XX}$ and $\lambda_S \equiv \sum_{i=1,2}\lambda_{\psi\bar{\psi}\chi h_i}$, then the cBE becomes
\begin{eqnarray}
\frac{dY_\psi}{dx} &=& - \lambda_{A1} \left(Y_\psi^2 - Y_{\psi,e}^2\right) -  \lambda_{S}\left(Y_\psi^2 - Y_\chi\frac{Y_{\psi ,e}^2}{Y_{\chi,e}}\right) \label{cBEappa}, 
 \\
 \frac{dY_\chi}{dx} &=& - \lambda_{A2} \left(Y_\chi^2 - Y_{\chi,e}^2\right) +  \frac{1}{2}\lambda_{S}\left(Y_\psi^2 - Y_\chi\frac{Y_{\psi ,e}^2}{Y_{\chi,e}}\right) .\label{cBEappb} 
\end{eqnarray}
First, we solve the cBE eq.~\ref{cBEappa} and \ref{cBEappb} numerically in Python with \texttt{solver$\_$ivp}, assuming thermal equilibrium at $x=1$. In order to solve the system, we assume $m_\psi = 100$ GeV, and $m_\chi = 150$ (solid red) and $m_\chi = 180$ GeV (dashed red). The results are shown in the plot in the left of Fig.~\ref{figapp}, where the solution for $Y_\chi$ results to be highly sensitive to $m_\chi$. Here, we are assuming $\lambda_{A1} = \lambda_{A2} = \lambda_{S}$, where each average annihilation cross section is taken to the value $10^{-9}$ GeV$^{-2}$. As it is clear, the heavier is $\chi$, the longer it stays in thermal equilibrium with the SM. This result has also been checked with \texttt{micrOMEGAs}. For comparison, in Fig.~\ref{figapp}(middle plot), we show the resulting behavior considering $\lambda_S = 0$, a typical freeze-out of two non-interacting DM particles. Thereby, the presence of the semi-annihilations in the cBE, results in a significant impact on $Y_\chi$, keeping $\chi$ longer in thermal equilibrium,  with a strong dependence on $m_\chi$.

In the following, we take a semi-analytical approach to solve the cBE. Based on what we have gotten in the numerical solution, we assume that $Y_\psi$ tracks its equilibrium yield for $x < 15$, which is our temperature region of interest, then we take $Y_\psi = Y_{\psi, e}$. Furthermore, following the \textit{freeze-out approximation} \cite{Cline:2013gha}, we take $Y_\chi \approx (1 + \delta)Y_{\chi,e}$, with $\delta$ a positive number that grows slowly. After some algebra, and taking $d\delta/dx \ll \delta$, we obtain
\begin{eqnarray}\label{foapp}
 -\frac{dY_{\chi e}}{dx}\bigg|_{x_f} = \lambda_{A2}(x_f)\frac{\delta_f(2 + \delta_f)}{1 + \delta_f}Y_{\chi e}^2(x_f) + \lambda_{S}(x_f)\frac{\delta_f}{1 + \delta_f}Y_{\psi,e}^2(x_f),
\end{eqnarray}
where we have evaluated all the quantities at $x_f$. This is a transcendental equation for $x_f$ which can be solved easily. We solve this equation considering $\delta_f = 1$\footnote{The solution to eq.~\ref{foapp} varies too slow with $\delta$, then it is safe to take the unit as a reference number.}, which is the moment at which starts the chemical decoupling. As it can be seen in Fig.~\ref{figapp}(right), the orange line represents the left side of eq.~\ref{foapp}, whereas the rest of the lines correspond to the r.h.s. of eq.~\ref{foapp}, with each line considering (w/s) and not (n/s) the semi-annihilation term proportional to $\lambda_S$. 

As it was anticipated by the numerical solution in the first part of this appendix, the overall effect of the presence of semi-annihilations in the cBE, makes $\chi$ to be longer in thermal contact with the plasma, since the contribution of the second term in the r.h.s of eq.~\ref{foapp} add a new positive contribution. Secondly, higher values of $m_\chi$ result in a later chemical freeze-out temperature in the case in which semi-annihilations are present (w/s case). Therefore, the numerical and the semi-analytical approach for the temperature decoupling of $Y_\chi$ agree, with the semi-annihilation not only predicting a bouncing as $m_\chi > m_\psi$, but impacting strongly the behavior of $Y_\chi$ before the bouncing.

\section{Cross sections}\label{Appendixcross}
In this appendix, we present algebraic expressions for each DM average annihilation cross section times relative velocity relevant for the calculation of indirect detection. All the cross sections here were obtained with \texttt{CalcHEP}. As we expand each cross-section in powers of $v^n$, with $v$ the relative velocity of the colliding non-relativistic DM particles, with $n$ taking positive even numbers, these expressions are not precise enough around poles or thresholds. In this way, we take the non-relativistic expansion $s=4m_i^2\left(1+\frac{v^2}{4}\right)$, with $i = \psi$ or $\chi$, retaining the $s$ and $p$ wave only. For the case of the fermion DM, we have
\begin{eqnarray}\small
    \braket{\sigma v}_{\psi\bar{\psi}\chi h_1} = \frac{g_\psi^4 \tan^2\theta\left(m_\chi^4 + 2m_\chi^2 m_{h_1}^2 + (m_{h_1}^2 - 4m_\psi^2)^2\right)^2\sqrt{m_\chi^4 - 2m_\chi^2 (m_{h_1}^2 + 4m_\psi^2) + (m_{h_1}^2 - 4m_\psi^2)^2}}{256 \pi m_\psi^4 (1 + \tan^2\theta)(m_\chi^2 - 4m_\psi^2)^2(m_\chi^2 + m_{h_1}^2 - 4m_\psi^2)^2}.
\end{eqnarray}
The expression for $\braket{\sigma v}_{\psi\bar{\psi}\chi h_2}$ is the same as the previous one but without $\tan^2\theta$ in the numerator, and $m_{h_1} \rightarrow m_{h_2}$. Additionally, the annihilation of the fermion DM into gauge bosons is velocity suppressed:
\begin{eqnarray}\small
    \braket{\sigma v}_{\psi\bar{\psi} W^+W^-} = \frac{g_\psi^2 m_Z^2 s_W^2 \tan\theta^2 \left(m_{h_1}^2-m_{h_2}^2\right)^2 \sqrt{1-\frac{m_W^2}{m_\psi^2}} \left(4 m_\psi^4-4 m_\psi^2 m_W^2+3 m_W^4\right)}{32 \text{pi} \left(\tan\theta^2+1\right)^2 v_h^2 \left(m_{h_1}^2-4 m_\psi^2\right)^2 \left(m_{h_2}^2-4 m_\psi^2\right)^2 \left(m_Z^2 - m_W^2\right)}v^2.
\end{eqnarray}
The annihilation cross-section for the pNGB are
\begin{eqnarray}\small
   \braket{\sigma v}_{\chi\chi W^+W^-} =  -\frac{g_\psi^2 m_\chi^2 m_Z^2 s_W^2 \tan^2\theta \sqrt{1-\frac{m_W^2}{m_\chi^2}} \left(4 m_\chi^4-4 m_\chi^2 m_W^2+3 m_W^4\right) \left(m_{h_1}^2- m_{h_2}^2\right)^2}{m_\psi^2 \pi \left(\tan^2\theta+1\right)^2 v_h^2 \left(m_{h_1}^2-4 m_\chi^2\right)^2 \left(m_{h_2}^2-4 m_\chi^2\right)^2 \left(m_W^2- m_Z^2\right)},
\end{eqnarray}
\begin{eqnarray}\small
   \braket{\sigma v}_{\chi\chi ZZ} =  \frac{g_\psi^2 m_\chi^2 m_Z^2 s_W^2 \tan^2\theta \sqrt{1-\frac{m_Z^2}{m_\chi^2}} \left(4 m_\chi^4-4 m_\chi^2 m_Z^2+3 m_Z^4\right) \left(m_{h_1}^2- m_{h_2}^2\right)^2}{2 m_\psi^2 \pi \left(\tan^2\theta+1\right)^2 v_h^2 \left(m_{h_1}^2-4 m_\chi^2\right)^2 \left(m_{h_2}^2-4 m_\chi^2\right)^2 \left(m_Z^2- m_W^2\right)},
\end{eqnarray}
The expressions for $\braket{\sigma v}_{\chi\chi h_ih_j}$, with $i,j=1,2$, are too long to be written here. However, they result to be $s$-wave, therefore relevant for indirect detection observables. 

\section{Scalar potential}\label{potentialap}
The complete scalar potential takes the form:
\begin{align}
	V(h_1, h_2, \chi) &= \frac{1}{2}m^2_1h^2_1 + \frac{1}{2}m^2_2h^2_2 + \frac{1}{2}m^2_\chi \chi^2 + \frac{m^2_2\cos\theta}{2v_s} h_2\chi^2 + \frac{m^2_1 + m^2_2 + (m^2_2 - m^2_1)\cos(2\theta)}{16v^2_s}\chi^4 \nonumber \\
	&\quad - \frac{m^2_1\sin\theta}{2v_s}h_1\chi^2 - \frac{[2(m^2_1 - m^2_2)v_s\cos^3\theta + v_h(-m^2_1 - m^2_2 + (m^2_1 - m^2_2)\cos(2\theta))\sin\theta]\sin\theta}{8v_hv^2_s}h_1^2\chi^2 \nonumber \\
	&\quad + \frac{m^2_1[3v_s\cos\theta + v_s\cos(3\theta) - 4v_h\sin^3\theta]}{8v_hv_s}h^3_1 + \frac{m^2_2[3v_h\cos\theta + v_h\cos(3\theta) + 4v_s\sin^3\theta]}{8v_hv_s}h^3_2 \nonumber \\
	&\quad - \frac{\cos\theta[v_h\cos\theta(-m^2_1 - m^2_2 + (m^2_1 - m^2_2)\cos(2\theta)) + 2(m^2_1 - m^2_2)v_s\sin^3\theta]}{8v_hv^2_s}h^2_2\chi^2 \nonumber \\
	&\quad + \frac{1}{16v^2_hv^2_s}\Bigl[-v^2_h\cos^4\theta[-m^2_1 - m^2_2 + (m^2_1 - m^2_2)\cos(2\theta)] - 4(m^2_1 - m^2_2)v_hv_s\cos^3\theta\sin^3\theta \nonumber \\
	&\quad + v_s^2[m^2_1 + m^2_2 + (m^2_1 - m^2_2)\cos(2\theta)]\sin^4\theta\Bigr]h^4_2 \nonumber \\
	&\quad + \frac{1}{16v^2_hv^2_s}\Bigl[v^2_s\cos^4\theta[m^2_1 + m^2_2 + (m^2_1 - m^2_2)\cos(2\theta)] - 4(m^2_1 - m^2_2)v_hv_s\cos^3\theta\sin^3\theta \nonumber \\
	&\quad + v_h^2[m^2_1 + m^2_2 + (-m^2_1 + m^2_2)\cos(2\theta)]\sin^4\theta\Bigr]h^4_1 \nonumber \\
	&\quad + \frac{(2m^2_1 + m_2^2)(v_s\cos\theta + v_h\sin\theta)\sin(2\theta)}{4v_hv_s}h^2_1h_2 + \frac{(m^2_1 + 2m^2_2)(-v_h\cos\theta + v_s\sin\theta)\sin(2\theta)}{4v_hv_s}h_1h^2_2 \nonumber \\
	&\quad + \frac{1}{16v^2_hv^2_s}\Bigl[v_s\cos\theta + v_h\sin\theta\Bigr]\Bigl[(3m^2_1 + m^2_2)v_s\cos\theta + (m^2_1 - m^2_2)v_s\cos(3\theta) \nonumber \\
	&\quad + 2v_h[-m^2_1 - m^2_2 + (m^2_1 - m^2_2)\cos(2\theta)]\sin\theta\Bigr]\sin(2\theta)h^3_1h_2 \nonumber \\
	&\quad + \frac{1}{16v^2_hv^2_s}\Bigl[v_h\cos\theta - v_s\sin\theta\Bigr]\Bigl[(m^2_1 + 3m^2_2)v_h\cos\theta - (m^2_1 - m^2_2)v_h\cos(3\theta) \nonumber \\
	&\quad + 2v_s[m^2_1 + m^2_2 + (m^2_1 - m^2_2)\cos(2\theta)]\sin\theta\Bigr]\sin(2\theta)h^3_2h_1 \nonumber \\
	&\quad + \frac{\cos\theta\sin\theta[-v_h(m^2_1 + m^2_2) + (m^2_1 - m^2_2)v_h\cos(2\theta) - (m^2_1 - m^2_2)v_s\sin(2\theta)]}{4v_hv^2_s}h_1h_2\chi^2 \nonumber \\
	&\quad + \frac{1}{64v^2_hv^2_s}\Bigl[-6(m^2_1 - m^2_2)v_hv_s\cos(4\theta) + 6(m^2_1 + m^2_2)(v^2_h + v^2_s)\sin(2\theta) \nonumber \\
	&\quad - (m^2_1 - m^2_2)[2v_hv_s + 3(v^2_h - v^2_s)\sin(4\theta)]\Bigr]\sin(2\theta)h^2_1h^2_2
\end{align}

\bibliography{bibliography}

\providecommand{\href}[2]{#2}\begingroup\raggedright\begin{thebibliography}{10}

\bibitem{Barger:2008jx}
V.~Barger, P.~Langacker, M.~McCaskey, M.~Ramsey-Musolf, and G.~Shaughnessy,
  ``{Complex Singlet Extension of the Standard Model},''
  \href{http://dx.doi.org/10.1103/PhysRevD.79.015018}{{\em Phys. Rev. D}
  {\bfseries 79} (2009) 015018},
  \href{http://arxiv.org/abs/0811.0393}{{\ttfamily arXiv:0811.0393 [hep-ph]}}.

\bibitem{Barger:2010yn}
V.~Barger, M.~McCaskey, and G.~Shaughnessy, ``{Complex Scalar Dark Matter
  vis-\textbackslash{}`{a}-vis CoGeNT, DAMA/LIBRA and XENON100},''
  \href{http://dx.doi.org/10.1103/PhysRevD.82.035019}{{\em Phys. Rev. D}
  {\bfseries 82} (2010) 035019},
  \href{http://arxiv.org/abs/1005.3328}{{\ttfamily arXiv:1005.3328 [hep-ph]}}.

\bibitem{Gross:2017dan}
C.~Gross, O.~Lebedev, and T.~Toma, ``{Cancellation Mechanism for
  Dark-Matter\textendash{}Nucleon Interaction},''
  \href{http://dx.doi.org/10.1103/PhysRevLett.119.191801}{{\em Phys. Rev.
  Lett.} {\bfseries 119} no.~19, (2017) 191801},
  \href{http://arxiv.org/abs/1708.02253}{{\ttfamily arXiv:1708.02253
  [hep-ph]}}.

\bibitem{Cline:2019okt}
J.~M. Cline and T.~Toma, ``{Pseudo-Goldstone dark matter confronts cosmic ray
  and collider anomalies},''
  \href{http://dx.doi.org/10.1103/PhysRevD.100.035023}{{\em Phys. Rev. D}
  {\bfseries 100} no.~3, (2019) 035023},
  \href{http://arxiv.org/abs/1906.02175}{{\ttfamily arXiv:1906.02175
  [hep-ph]}}.

\bibitem{Feng:2003xh}
J.~L. Feng, A.~Rajaraman, and F.~Takayama, ``{Superweakly interacting massive
  particles},'' \href{http://dx.doi.org/10.1103/PhysRevLett.91.011302}{{\em
  Phys. Rev. Lett.} {\bfseries 91} (2003) 011302},
  \href{http://arxiv.org/abs/hep-ph/0302215}{{\ttfamily arXiv:hep-ph/0302215}}.

\bibitem{Bringmann:2021tjr}
T.~Bringmann, P.~F. Depta, M.~Hufnagel, J.~T. Ruderman, and K.~Schmidt-Hoberg,
  ``{Dark Matter from Exponential Growth},''
  \href{http://dx.doi.org/10.1103/PhysRevLett.127.191802}{{\em Phys. Rev.
  Lett.} {\bfseries 127} no.~19, (2021) 191802},
  \href{http://arxiv.org/abs/2103.16572}{{\ttfamily arXiv:2103.16572
  [hep-ph]}}.

\bibitem{Puetter:2022ucx}
L.~Puetter, J.~T. Ruderman, E.~Salvioni, and B.~Shakya, ``{Bouncing Dark
  Matter},'' \href{http://arxiv.org/abs/2208.08453}{{\ttfamily arXiv:2208.08453
  [hep-ph]}}.

\bibitem{Katz:2020ywn}
A.~Katz, E.~Salvioni, and B.~Shakya, ``{Split SIMPs with Decays},''
  \href{http://dx.doi.org/10.1007/JHEP10(2020)049}{{\em JHEP} {\bfseries 10}
  (2020) 049}, \href{http://arxiv.org/abs/2006.15148}{{\ttfamily
  arXiv:2006.15148 [hep-ph]}}.

\bibitem{Ho:2022erb}
S.-Y. Ho, P.~Ko, and C.-T. Lu, ``{Scalar and fermion two-component SIMP dark
  matter with an accidental $\mathbb Z_{4}$ symmetry},''
  \href{http://dx.doi.org/10.1007/JHEP03(2022)005}{{\em JHEP} {\bfseries 03}
  (2022) 005}, \href{http://arxiv.org/abs/2201.06856}{{\ttfamily
  arXiv:2201.06856 [hep-ph]}}.

\bibitem{Ho:2022tbw}
S.-Y. Ho, ``{An asymmetric SIMP dark matter model},''
  \href{http://dx.doi.org/10.1007/JHEP10(2022)182}{{\em JHEP} {\bfseries 10}
  (2022) 182}, \href{http://arxiv.org/abs/2207.13373}{{\ttfamily
  arXiv:2207.13373 [hep-ph]}}.

\bibitem{Ghosh:2023dhj}
P.~Ghosh and S.~Jeesun, ``{Reviving sub-TeV $SU(2)_L$ lepton doublet dark
  matter},'' \href{http://dx.doi.org/10.1140/epjc/s10052-023-12039-z}{{\em Eur.
  Phys. J. C} {\bfseries 83} no.~9, (2023) 880},
  \href{http://arxiv.org/abs/2306.12906}{{\ttfamily arXiv:2306.12906
  [hep-ph]}}.

\bibitem{Weinberg:2013kea}
S.~Weinberg, ``{Goldstone Bosons as Fractional Cosmic Neutrinos},''
  \href{http://dx.doi.org/10.1103/PhysRevLett.110.241301}{{\em Phys. Rev.
  Lett.} {\bfseries 110} no.~24, (2013) 241301},
  \href{http://arxiv.org/abs/1305.1971}{{\ttfamily arXiv:1305.1971
  [astro-ph.CO]}}.

\bibitem{Garcia-Cely:2013wda}
C.~Garcia-Cely, A.~Ibarra, and E.~Molinaro, ``{Cosmological and astrophysical
  signatures of dark matter annihilations into pseudo-Goldstone bosons},''
  \href{http://dx.doi.org/10.1088/1475-7516/2014/02/032}{{\em JCAP} {\bfseries
  02} (2014) 032}, \href{http://arxiv.org/abs/1312.3578}{{\ttfamily
  arXiv:1312.3578 [hep-ph]}}.

\bibitem{Arcadi:2017kky}
G.~Arcadi, M.~Dutra, P.~Ghosh, M.~Lindner, Y.~Mambrini, M.~Pierre, S.~Profumo,
  and F.~S. Queiroz, ``{The waning of the WIMP? A review of models, searches,
  and constraints},''
  \href{http://dx.doi.org/10.1140/epjc/s10052-018-5662-y}{{\em Eur. Phys. J. C}
  {\bfseries 78} no.~3, (2018) 203},
  \href{http://arxiv.org/abs/1703.07364}{{\ttfamily arXiv:1703.07364
  [hep-ph]}}.

\bibitem{Yaguna:2021rds}
C.~E. Yaguna and O.~Zapata, ``{Fermion and scalar two-component dark matter
  from a Z4 symmetry},''
  \href{http://dx.doi.org/10.1103/PhysRevD.105.095026}{{\em Phys. Rev. D}
  {\bfseries 105} no.~9, (2022) 095026},
  \href{http://arxiv.org/abs/2112.07020}{{\ttfamily arXiv:2112.07020
  [hep-ph]}}.

\bibitem{Pokorski:2021qgt}
S.~Pokorski and K.~Sakurai, ``{Goldstone boson decays and chiral anomalies},''
\newblock 5, 2021.
\newblock \href{http://arxiv.org/abs/2105.04877}{{\ttfamily arXiv:2105.04877
  [hep-ph]}}.

\bibitem{Azevedo:2018oxv}
D.~Azevedo, M.~Duch, B.~Grzadkowski, D.~Huang, M.~Iglicki, and R.~Santos,
  ``{Testing scalar versus vector dark matter},''
  \href{http://dx.doi.org/10.1103/PhysRevD.99.015017}{{\em Phys. Rev. D}
  {\bfseries 99} no.~1, (2019) 015017},
  \href{http://arxiv.org/abs/1808.01598}{{\ttfamily arXiv:1808.01598
  [hep-ph]}}.

\bibitem{DiazSaez:2021pmg}
B.~D\'\i{}az~S\'aez, P.~Escalona, S.~Norero, and A.~R. Zerwekh, ``{Fermion
  singlet dark matter in a pseudoscalar dark matter portal},''
  \href{http://dx.doi.org/10.1007/JHEP10(2021)233}{{\em JHEP} {\bfseries 10}
  (2021) 233}, \href{http://arxiv.org/abs/2105.04255}{{\ttfamily
  arXiv:2105.04255 [hep-ph]}}.

\bibitem{Belyaev:2022qnf}
A.~Belyaev, G.~Cacciapaglia, D.~Locke, and A.~Pukhov, ``{Minimal Consistent
  Dark Matter models for systematic experimental characterisation: Fermion Dark
  Matter},'' \href{http://arxiv.org/abs/2203.03660}{{\ttfamily arXiv:2203.03660
  [hep-ph]}}.

\bibitem{Belanger:2014vza}
G.~B\'elanger, F.~Boudjema, A.~Pukhov, and A.~Semenov, ``{micrOMEGAs4.1: two
  dark matter candidates},''
  \href{http://dx.doi.org/10.1016/j.cpc.2015.03.003}{{\em Comput. Phys.
  Commun.} {\bfseries 192} (2015) 322--329},
  \href{http://arxiv.org/abs/1407.6129}{{\ttfamily arXiv:1407.6129 [hep-ph]}}.

\bibitem{Semenov:2002jw}
A.~V. Semenov, ``{LanHEP: A Package for automatic generation of Feynman rules
  in field theory. Version 2.0},''
  \href{http://arxiv.org/abs/hep-ph/0208011}{{\ttfamily arXiv:hep-ph/0208011}}.

\bibitem{Belanger:2013ywg}
G.~Belanger, F.~Boudjema, and A.~Pukhov,
  \href{http://dx.doi.org/10.1142/9789814390163_0012}{``{micrOMEGAs : a code
  for the calculation of Dark Matter properties in generic models of particle
  interaction},''} in {\em {Theoretical Advanced Study Institute in Elementary
  Particle Physics}: {The Dark Secrets of the Terascale}}, pp.~739--790.
\newblock 2013.
\newblock \href{http://arxiv.org/abs/1402.0787}{{\ttfamily arXiv:1402.0787
  [hep-ph]}}.

\bibitem{Graesser:2011wi}
M.~L. Graesser, I.~M. Shoemaker, and L.~Vecchi, ``{Asymmetric WIMP dark
  matter},'' \href{http://dx.doi.org/10.1007/JHEP10(2011)110}{{\em JHEP}
  {\bfseries 10} (2011) 110}, \href{http://arxiv.org/abs/1103.2771}{{\ttfamily
  arXiv:1103.2771 [hep-ph]}}.

\bibitem{Krnjaic:2015mbs}
G.~Krnjaic, ``{Probing Light Thermal Dark-Matter With a Higgs Portal
  Mediator},'' \href{http://dx.doi.org/10.1103/PhysRevD.94.073009}{{\em Phys.
  Rev. D} {\bfseries 94} no.~7, (2016) 073009},
  \href{http://arxiv.org/abs/1512.04119}{{\ttfamily arXiv:1512.04119
  [hep-ph]}}.

\bibitem{Yang:2022zlh}
K.-C. Yang, ``{Freeze-out forbidden dark matter in the hidden sector in the
  mass range from sub-GeV to TeV},''
  \href{http://dx.doi.org/10.1007/JHEP11(2022)083}{{\em JHEP} {\bfseries 11}
  (2022) 083}, \href{http://arxiv.org/abs/2209.10827}{{\ttfamily
  arXiv:2209.10827 [hep-ph]}}.

\bibitem{Amiri:2022cbv}
A.~Amiri, B.~D\'\i{}az~S\'aez, and K.~Ghorbani, ``{(sub)GeV Dark Matter in the
  $U(1)_X$ Higgs Portal Model},''
  \href{http://dx.doi.org/10.1016/j.physletb.2023.138119}{{\em Phys. Lett. B}
  {\bfseries 844} (2023) 138119},
  \href{http://arxiv.org/abs/2209.11723}{{\ttfamily arXiv:2209.11723
  [hep-ph]}}.

\bibitem{planck2018}
{{Planck collaboration}, N. Aghanim et al.}, ``{Planck 2018 results. VI.
  Cosmological parameters},''  (2018) ,
\href{http://arxiv.org/abs/1807.06209}{{\ttfamily arXiv:1807.06209
  [astro-ph.CO]}}.

\bibitem{Alanne:2020jwx}
T.~Alanne, N.~Benincasa, M.~Heikinheimo, K.~Kannike, V.~Keus, N.~Koivunen, and
  K.~Tuominen, ``{Pseudo-Goldstone dark matter: gravitational waves and
  direct-detection blind spots},''
  \href{http://dx.doi.org/10.1007/JHEP10(2020)080}{{\em JHEP} {\bfseries 10}
  (2020) 080}, \href{http://arxiv.org/abs/2008.09605}{{\ttfamily
  arXiv:2008.09605 [hep-ph]}}.

\bibitem{Ishiwata:2018sdi}
K.~Ishiwata and T.~Toma, ``{Probing pseudo Nambu-Goldstone boson dark matter at
  loop level},'' \href{http://dx.doi.org/10.1007/JHEP12(2018)089}{{\em JHEP}
  {\bfseries 12} (2018) 089}, \href{http://arxiv.org/abs/1810.08139}{{\ttfamily
  arXiv:1810.08139 [hep-ph]}}.

\bibitem{LZ:2022ufs}
{\bfseries LZ} Collaboration, J.~Aalbers {\em et~al.}, ``{First Dark Matter
  Search Results from the LUX-ZEPLIN (LZ) Experiment},''
  \href{http://arxiv.org/abs/2207.03764}{{\ttfamily arXiv:2207.03764
  [hep-ex]}}.

\bibitem{Aprile_2018}
J.~e.~a. Aprile, E.~Aalbers, ``Dark matter search results from a one ton-year
  exposure of xenon1t,''
  \href{http://dx.doi.org/10.1103/physrevlett.121.111302}{{\em Physical Review
  Letters} {\bfseries 121} no.~11, (Sep, 2018) }.
  \url{http://dx.doi.org/10.1103/PhysRevLett.121.111302}.

\bibitem{XENON:2020kmp}
{\bfseries XENON} Collaboration, E.~Aprile {\em et~al.}, ``{Projected WIMP
  sensitivity of the XENONnT dark matter experiment},''
  \href{http://dx.doi.org/10.1088/1475-7516/2020/11/031}{{\em JCAP} {\bfseries
  11} (2020) 031}, \href{http://arxiv.org/abs/2007.08796}{{\ttfamily
  arXiv:2007.08796 [physics.ins-det]}}.

\bibitem{Falkowski:2015iwa}
A.~Falkowski, C.~Gross, and O.~Lebedev, ``{A second Higgs from the Higgs
  portal},'' \href{http://dx.doi.org/10.1007/JHEP05(2015)057}{{\em JHEP}
  {\bfseries 05} (2015) 057}, \href{http://arxiv.org/abs/1502.01361}{{\ttfamily
  arXiv:1502.01361 [hep-ph]}}.

\bibitem{Arcadi:2016qoz}
G.~Arcadi, C.~Gross, O.~Lebedev, S.~Pokorski, and T.~Toma, ``{Evading Direct
  Dark Matter Detection in Higgs Portal Models},''
  \href{http://dx.doi.org/10.1016/j.physletb.2017.03.044}{{\em Phys. Lett. B}
  {\bfseries 769} (2017) 129--133},
  \href{http://arxiv.org/abs/1611.09675}{{\ttfamily arXiv:1611.09675
  [hep-ph]}}.

\bibitem{ATLAS:2022yvh}
{\bfseries ATLAS} Collaboration, G.~Aad {\em et~al.}, ``{Search for invisible
  Higgs-boson decays in events with vector-boson fusion signatures using 139
  fb$^{-1}$ of proton-proton data recorded by the ATLAS experiment},''
  \href{http://dx.doi.org/10.1007/JHEP08(2022)104}{{\em JHEP} {\bfseries 08}
  (2022) 104}, \href{http://arxiv.org/abs/2202.07953}{{\ttfamily
  arXiv:2202.07953 [hep-ex]}}.

\bibitem{Bonilla:2023egs}
C.~Bonilla, A.~E. C\'arcamo~Hern\'andez, B.~D\'\i{}az~S\'aez, S.~Kovalenko, and
  J.~M. Gonz\'alez, ``{Dark Matter from a Radiative Inverse Seesaw Majoron
  Model},'' \href{http://arxiv.org/abs/2306.08453}{{\ttfamily arXiv:2306.08453
  [hep-ph]}}.

\bibitem{Cline:2013gha}
J.~M. Cline, K.~Kainulainen, P.~Scott, and C.~Weniger, ``{Update on scalar
  singlet dark matter},''
  \href{http://dx.doi.org/10.1103/PhysRevD.88.055025}{{\em Phys. Rev. D}
  {\bfseries 88} (2013) 055025},
  \href{http://arxiv.org/abs/1306.4710}{{\ttfamily arXiv:1306.4710 [hep-ph]}}.
  [Erratum: Phys.Rev.D 92, 039906 (2015)].

\end{thebibliography}\endgroup
\bibliographystyle{utphys}

\end{document}